\documentclass[12pt]{article}
\usepackage{amsmath} 
\input{epsf} 
\def\gtap{\raisebox{-.6ex}{\rlap{$\,\sim\,$}} \raisebox{.4ex}{$\,>\,$}} 
 
\def\naive{na\"{\i}ve} 
\newcommand\as{\alpha_{\mathrm{S}}} 
\newcommand\f[2]{\frac{#1}{#2}}

\def\beq{\begin{equation}} 
\def\eeq{\end{equation}} 
\def\beeq{\begin{eqnarray}} 
\def\eeeq{\end{eqnarray}} 
 
\def\to{\rightarrow}
 
\def\nn{\nonumber} 
 
\def\qt{q_T}
\def\bqt{{\bf q_T}}

\def\ms{${\overline {\rm MS}}$}

\def\b0{b_0}

\def\vep{\varepsilon}

\begin{document} 

\begin{titlepage}
\renewcommand{\thefootnote}{\fnsymbol{footnote}}
\par \vspace*{10mm}

\begin{center}
{\Large \bf
QCD transverse-momentum resummation \\
\vskip .3cm
in gluon fusion processes}
\end{center}
\par \vspace{2mm}
\begin{center}
{\bf Stefano Catani} and 
{\bf Massimiliano Grazzini}

\vspace{5mm}

INFN, Sezione di Firenze and Dipartimento di Fisica,
Universit\`a di Firenze,\\ 
I-50019 Sesto Fiorentino, Florence, Italy

\vspace{5mm}

\end{center}

\par \vspace{2mm}
\begin{center} {\large \bf Abstract} \end{center}
\begin{quote}
\pretolerance 10000

We consider the production of a generic
system of non-strongly
interacting particles with a high total invariant mass $M$
in hadron collisions.
We examine the transverse-momentum ($q_T$) distribution of the system 
in the small-$q_T$ region ($q_T \ll M$),
and we present a study of the perturbative QCD contributions that
are enhanced by powers of large logarithmic terms of the type $\ln (M^2/\qt^2)$.
These terms can be resummed to all orders in QCD perturbation theory.
The partonic production mechanism of the final-state system  
can be controlled by quark--antiquark $(q{\bar q})$ annihilation and/or by 
gluon fusion. 
The resummation formalism for the 
$q{\bar q}$
annihilation subprocess
is well established, and it is usually extrapolated to the gluon fusion 
subprocess. We point out that this \naive\ extrapolation is not correct,
and we present the all-order  resummation formula for the $q_T$   
distribution in gluon fusion processes. The gluon fusion resummation formula
has a richer structure than the resummation formula in $q{\bar q}$
annihilation. The additional structure originates from
collinear correlations that are a specific feature of
the evolution of the colliding hadrons into gluon partonic states.
In the $\qt$ cross section at small values of $\qt$, 
these gluon collinear correlations produce coherent spin correlations 
between the helicity states of the initial-state gluons
and definite azimuthal-angle correlations between the final-state
particles of the observed high-mass system.

\end{quote}

\vspace*{\fill}
\begin{flushleft}
November 2010
\end{flushleft}
\end{titlepage}

\setcounter{footnote}{1}
\renewcommand{\thefootnote}{\fnsymbol{footnote}}

\section{Introduction}
\label{sec:intro}

The properties of the transverse-momentum distributions of systems of high
invariant mass that are
produced at high-energy hadron colliders are important for QCD and electroweak
studies
and for physics studies beyond the Standard Model (SM).

We consider the inclusive hard-scattering process
\begin{equation}
\label{process}
h_1(p_1) + h_2(p_2) \to F(\{q_i\}) + X \;\;,
\end{equation}
where the collision of the two hadrons $h_1$ and $h_2$ with
momenta $p_1$ and $p_2$ produces the triggered final-state system $F$,
accompanied by an arbitrary and undetected final state $X$.
We denote by $\sqrt s$ the centre--of--mass energy of the colliding hadrons,
which are treated as massless particles
$(s= (p_1+p_2)^2 = 2p_1p_2)$.
The observed final state $F$ is a generic system of {\em one} or {\em more}
particles with momenta $q_i^{\mu}$ $(i=3,4,5,\dots)$. The total momentum 
of $F$ is
denoted by $q^{\mu}$ ($q=\sum_i q_i$), 
and it can be expressed in terms of
the total invariant mass $M$ $(q^2=M^2)$, the transverse momentum $\bf \qt$
with respect to the direction of the colliding hadrons,
and the rapidity $y$ $(2y = \ln (p_2q/p_1q))$ in the centre--of--mass
system of the collision.
Throughout the paper, we limit ourselves
to considering the case in which the system $F$
is formed by non-strongly interacting particles, 
such as vector bosons $(\gamma, W, Z, \dots)$, Drell--Yan (DY) lepton pairs,
Higgs particles and so forth.

Provided the invariant mass $M$ is large
$(M \gg \Lambda_{QCD}, \;\Lambda_{QCD}$ being the QCD scale), the production
cross section and associated kinematical distributions 
of the process in Eq.~(\ref{process}), can be evaluated by using QCD
perturbation theory. The cross section is expressed as a convolution of 
partonic
cross sections $d{\hat \sigma}_{ab}$ $(a,b=q, {\bar q},g)$ with the parton
densities of the colliding hadrons. The partonic
cross section $d{\hat \sigma}_{ab}$ is computed as a power series expansion in
the QCD coupling $\as(M^2)$ by considering the corresponding partonic
subprocess $a + b \to F + \dots$, where the dots denote final-state partons.
Since $F$ is a system of colourless particles, the partonic
subprocesses include quark-antiquark $(q{\bar q})$ annihilation,
\begin{equation}
\label{ann}
q + {\bar q} \to F \;\;,
\end{equation}
and gluon fusion,
\beq
\label{fus}
g + g \to F \;\;.
\eeq

In this paper we are interested in considering the process of 
Eq.~(\ref{process}) in kinematical configurations where the transverse 
momentum $\qt$ of the system $F$ is small (say, $\qt \ll M$).
Unless the subprocesses in Eqs.~(\ref{ann}) and (\ref{fus}) are
both forbidden by selection rules related to the nature of $F$
(e.g., $gg \to W^\pm$ is forbidden), these subprocesses produce the system 
$F$  with $\qt = 0$. The system $F$ acquires a non-vanishing transverse
momentum through higher-order QCD radiative corrections to the 
subprocesses in Eqs.~(\ref{ann}) and (\ref{fus}). Nonetheless,
the {\em bulk} of the events is still produced in the small-$\qt$ region.

The perturbative-QCD computation of the partonic
cross sections $d{\hat \sigma}_{ab}$ in powers of $\as(M^2)$
shows that high-order coefficients contain logarithmic terms
of the type $\ln^m (M^2/q_T^2)$. Although $\as(M^2)$ is small, these
logarithmic terms can be large in the small-$\qt$ region ($\qt \ll M$),
thus spoiling the quantitative convergence of the expansion in powers of
$\as(M^2)$ 
(at each
fixed order in $\as$, the partonic cross
section eventually diverges to either $\,+\infty$ or $\,-\infty$ 
by considering the
limit $\qt \to 0$).
To obtain reliable perturbative predictions in the small-$\qt$ region, the
logarithmically-enhanced terms
have to be evaluated at sufficiently-high perturbative 
orders\footnote{The `sufficiently-high' order depends on the specific
$\qt$ region of interest in each specific process; this order cannot be
specified `a priori'.},
and possibly {\em resummed} to all orders in $\as(M^2)$.

The small-$\qt$ logarithmic terms have their physical origin from multiple 
radiation of final-state partons that are soft 
and/or collinear to the colliding hadrons (partons).
The method and the formalism to resum the logarithmically-enhanced terms
at small $q_T$ was developed in the eighties
[\ref{Dokshitzer:hw}--\ref{Catani:vd}].
Subsequent, and important, theoretical progress in this field regards,
for instance, the explicit computation of high-order resummation coefficients
[\ref{deFlorian:2000pr}--\ref{Becher:2010tm}]
(see additional comments in Sect.~\ref{sec:revqqbar}) 
and the understanding of their universality (process-independent) structure
[\ref{deFlorian:2000pr}, \ref{Catani:2000vq}].

In this paper we consider small-$\qt$ resummation, and we deal with an issue
that is also related  to universality. The issue regards the relation between
processes that are controlled by $q{\bar q}$ annihilation and by gluon fusion.
Transverse-momentum resummation was originally worked out for the DY process
[\ref{Dokshitzer:hw}--\ref{Collins:1984kg}], which is driven by   
$q{\bar q}$ annihilation. The resummation structure that emerges in the 
DY process was then {\em customarily} used 
(see, e.g., Refs.~[\ref{Balazs:2000wv}--\ref{Dreiner:2006sv}] 
and references therein) 
for many other processes of the class in Eq.~(\ref{process}).
Such processes include, for instance, the production of the SM Higgs boson
[\ref{Balazs:2000wv}--\ref{Bozzi:2007pn}], of photon pairs
[\ref{Balazs:2006cc}], of vector boson pairs such as  $ZZ$ [\ref{ZZ}]
and $W^+W^-$ [\ref{Grazzini:2005vw}],
and of slepton pairs
[\ref{Bozzi:2006fw}, \ref{Dreiner:2006sv}].
In particular, Higgs boson production is driven by gluon fusion, whereas 
diphoton and diboson production receives contributions from both the
$q{\bar q}$ annihilation and gluon fusion subprocesses.

In the present contribution, we point out that there are some key differences 
between $q{\bar q}$ annihilation and gluon fusion. These differences
have escaped detection until recent findings
[\ref{Catani:2007vq}, \ref{Nadolsky:2007ba}, \ref{Mantry:2009qz}].
The physical origin of the differences is due to specific
{\em collinear correlations} (see Sect.~\ref{sec:ggfus}) 
that are a distinctive feature of
the perturbative evolution of 
the colliding hadrons into gluon initial states.
Analogous correlations are not produced by the perturbative evolution of 
spin unpolarized hadrons into quark or antiquark initial states. 

As a consequence of these differences, transverse-momentum resummation in
gluon fusion subprocesses has a 'richer' structure than in 
$q{\bar q}$ annihilation subprocesses. The small-$\qt$ resummation
formalism for the DY process [\ref{Dokshitzer:hw}--\ref{Collins:1984kg}]
has to be modified and extended to deal with 
gluon fusion subprocesses. 
In particular, in gluon fusion subprocesses, 
gluon collinear correlations produce 
{\em spin} and {\em azimuthal} correlations that are logarithmically enhanced
in the small-$\qt$ region.

In the following 
we present and discuss our main general 
results on transverse-momentum resummation 
in gluon fusion processes. Then, we shall comment on 
Refs.~[\ref{Catani:2007vq}, \ref{Nadolsky:2007ba}, \ref{Mantry:2009qz}].
Details about the derivation of our results, and the illustration of further 
related results, will appear elsewhere  [\ref{CGinprep}].

The outline of the paper is as follows.
In Sect.~\ref{sec:revqqbar}, we briefly review the classical QCD results on
transverse-momentum resummation. These results, which mostly derive from 
studies of the DY process, are presented using a general and process-independent
notation that is useful for the subsequent presentation of $\qt$ resummation in
gluon fusion processes.
In Sect.~\ref{sec:ggfus}, we present our all-order resummation formula for generic
transverse-momentum cross sections controlled by gluon fusion. We illustrate the
structure of the resummation formula, and we discuss its origin from
quantum-mechanical correlations (interference effects) produced by the
collinear-parton radiation that accompanies the gluon fusion hard-scattering
subprocess. We also explicitly consider the specific example of SM Higgs boson
production.
In Sect.~\ref{sec:hel}, we reformulate
$\qt$ resummation in the helicity space
of the colliding gluons. We show how gluon collinear correlations are related to
helicity-flip phenomena in the hard-scattering subprocess.
In Sect.~\ref{sec:azav}, we specify the gluon fusion resummation formula
for azimuthally-averaged transverse-momentum cross sections. 
We point out that the differences between the $q{\bar q}$ annihilation and gluon
fusion channels persist even after having performed the integration over the
azimuthal angle of the transverse-momentum vector.
Section~\ref{sec:FB} is devoted to derive and discuss the general structure of the
azimuthal-angle correlations embodied in the gluon fusion
resummation formula.
Few summarizing remarks are presented in Sect.~\ref{sec:sum}.

\section{Small-$\qt$ resummation in impact parameter space}
\label{sec:revqqbar}

In this section we recall the `classical' formalism
[\ref{Dokshitzer:hw}--\ref{Catani:vd}, \ref{Catani:2000vq}]
of transverse-momentum resummation in impact parameter space.
This illustration sets the stage for the presentation of our 
results 
on small-$\qt$ resummation in gluon fusion processes 
(see Sects.~\ref{sec:ggfus}--\ref{sec:FB}).

We consider the process in Eq.~(\ref{process}), and we introduce the 
corresponding multidifferential cross section
\beq
\label{diffxs}
\f{d\sigma_F}{d^2{\bqt} \;dM^2 \;dy \;d{\bf\Omega}} 
\,(p_1, p_2;\bqt,M,y,
{\bf\Omega} )
\;\;.
\eeq
The differential cross section depends on the total momentum of the system $F$
(i.e. on the variables $\bqt, M, y$) and, to be quite general, it can also  
depend on additional variables that specify the kinematics of the particles 
in the system $F$. In Eq.~(\ref{diffxs}) these additional variables
are generically denoted as ${\bf\Omega}= \{\Omega_A,\Omega_B, \dots \}$
(correspondingly, we define $d{\bf\Omega} \equiv d\Omega_A d\Omega_B \dots)$. 
They can be, for instance, 
the rapidity $y_i$ and the azimuthal angle $\phi(\bqt_i)$ of one of the
particles (with momentum $q_i$) in the system $F$.
In general, we only assume that the kinematical variables 
$\{\Omega_A,\Omega_B, \dots \}$ are independent of $\bqt, M$ and $y$.

Considering the $\bqt$ dependence of the multidifferential cross section
in Eq.~(\ref{diffxs})
within perturbative QCD, we introduce the following
decomposition:
\beq
\label{Fdec}
d\sigma_F =
d\sigma_F^{({\rm sing})} +
\; d\sigma_F^{({\rm reg})}
\;\;.
\eeq
Both terms on the right-hand side are obtained as convolutions of
partonic cross sections and the scale-dependent parton distributions 
$f_{a/h}(x,\mu^2)$  ($a=q_f, {\bar q}_f, g$ is the parton label) of the
colliding hadrons\footnote{Throughout the paper we always use
parton densities as defined in the \ms\ factorization scheme, 
and $\as(q^2)$ is
the QCD running coupling in the \ms\ renormalization scheme.}. 
The distinction between the two terms is purely theoretical.
The partonic cross sections that enter the singular component (the first term
on the right-hand side) contain all the contributions that are enhanced
(or `singular') at small $\qt$. These contributions are proportional to
$\delta^{(2)}(\bqt)$ or to large logarithms\footnote{To be precise,
the logarithms are combined with corresponding 'contact' terms, which are
proportional to $\delta^{(2)}(\bqt)$. These combinations define regularized 
(integrable) `plus
distributions' $\left[\f{1}{\qt^2}\ln^m (M^2/\qt^2)\right]_+$ with respect to
$\bqt$.} of the type 
$\f{1}{\qt^2}\ln^m (M^2/\qt^2)$.
On the
contrary, the partonic cross sections of the second term on the right-hand side
are regular (i.e. free of logarithmic terms)
order-by-order in perturbation theory as $\qt \to 0$. To be precise, the
integration of $d\sigma_F^{({\rm reg})}/d^2{\bqt}$ over the range 
$0 \leq \qt \leq Q_0$ leads to a finite result that, at each
fixed order in $\as$, {\em vanishes} in the limit $Q_0 \to 0$.
 
The regular component $d\sigma_F^{({\rm reg})}$
of the $\qt$
cross section
is definitely process dependent. In this paper we limit ourselves to considering 
the singular component, which has a universal (process-independent)
structure. 

To simplify the presentation of the all-order (resummed) structure of the
singular component of the $\bqt$ differential cross section in 
Eq.~(\ref{diffxs}), we introduce a shorthand (symbolical) notation in several
places. For instance, the singular component of Eq.~(\ref{diffxs}) is simply
denoted by $\left[ d\sigma_F \right]$; namely, we define
\beq
\label{singsig}
\left[ d\sigma_F \right] \equiv \f{d\sigma_F^{({\rm sing})}}{d^2{\bqt} 
\;dM^2 \;dy \;d{\bf\Omega}} 
\,(p_1, p_2;\bqt,M,y,
{\bf\Omega} )
\;\;.
\eeq
The transverse-momentum resummation formula can be 
written in the following
factorized form [\ref{Collins:1984kg}, \ref{Catani:2000vq}]:
\beeq
\label{qtycross}
&&\left[ d\sigma_F \right] =\f{M^2}{s}
\sum_{c=q, {\bar q},g} \left[ d\sigma_{c{\bar c}, \,F}^{(0)} \right]
\int \f{d^2{\bf b}}{(2\pi)^2} \;\, e^{i {\bf b}\cdot \bqt} \;
  S_c(M,b)\nn \\
&& \;\;\;\; \times \;
\sum_{a_1,a_2} \,
\int_{x_1}^1 \f{dz_1}{z_1} \,\int_{x_2}^1 \f{dz_2}{z_2} 
\; \left[ H^F C_1 C_2 \right]_{c{\bar c};a_1a_2}
\;f_{a_1/h_1}(x_1/z_1,b_0^2/b^2)
\;f_{a_2/h_2}(x_2/z_2,b_0^2/b^2) \;
\;, 
\eeeq
where $b_0=2e^{-\gamma_E}$
($\gamma_E=0.5772\dots$ is the Euler number) is a numerical coefficient,
and the kinematical variables $x_1$ and $x_2$ are 
\beq
\label{xo}
x_1= \f{M}{\sqrt s} \;e^{+y} \;\;, \quad \quad
x_2=\f{M}{\sqrt s} \;e^{-y} \;\;.
\eeq
The function $S_c(M,b)$ and the functions symbolically denoted by
$\left[ d\sigma^{(0)}_F \right]$ and $\left[ H^F C_1 C_2 \right]$ are specified
below.

The right-hand side of Eq.~(\ref{qtycross}) involves the Fourier transformation
with respect to the
{\em impact parameter} ${\bf b}$ and two convolutions over the 
longitudinal-momentum fractions $z_1$ and $z_2$.
The parton densities 
$f_{a_i/h_i}(x,\mu^2)$ of the colliding hadrons are evaluated at the scale
$\mu= b_0/b$, which depends on the impact parameter.

We note that in the context of the study of the present paper, the resummation
formula in Eq.~(\ref{qtycross}), and the resummation formulae in 
Sects.~\ref{sec:ggfus}--\ref{sec:azav}, have a purely perturbative-QCD 
content\footnote{Throughout the paper we
do not consider the inclusion of any non-perturbative contributions, such as, 
for instance, those
first introduced in Ref.~[\ref{Collins:va}].}, analogously to customary
fixed-order calculations of hard-scattering cross sections in hadron collisions.
Using the Altarelli--Parisi evolution equations, the parton densities 
$f_{a/h}(x,b_0^2/b^2)$ can be expressed [\ref{Catani:2000vq}] in terms of the
corresponding parton densities $f_{a/h}(x,\mu_F^2)$ at the evolution scale
$\mu= \mu_F$, where $\mu_F$ is the customary {\em factorization scale}
that enters fixed-order calculations. Having done that, all the remaining factors
on the right-hand side of Eq.~(\ref{qtycross}) are partonic contributions that
can be expanded in powers of $\as(M^2)$ at arbitrary perturbative orders.

The small-$\qt$ region where $\qt \ll M$ corresponds in impact parameter space
to the large-$b$ region where $b \gg 1/M$. The perturbative expansion of the
$\bf b$ space integrand in Eq.~(\ref{qtycross}) produces large perturbative
coefficients of the type $\ln^m (b^2M^2)$: these coefficients lead to the 
small-$\qt$ logarithmic terms 
$\left[\f{1}{\qt^2}\ln^{m-1} (M^2/\qt^2)\right]_+$, through the evaluation
of the Fourier transformation from $\bf b$ space to $\bqt$ space.

The factor 
$\left[ d\sigma_{c{\bar c}, \,F}^{(0)} \right]$
in Eq.~(\ref{qtycross}) depends on the process (i.e. on the specific final 
state system $F$ and its kinematics). This factor is the Born level cross
section $d{\hat \sigma}^{(0)}$ (i.e. the cross section at its corresponding {\em lowest
order} in $\as$) of the partonic subprocesses $c + {\bar c} \to F$
in Eqs.~(\ref{ann}) and (\ref{fus}). Making the symbolic notation explicit, we
write:
\beq
\label{sig0}
\left[ d\sigma_{c{\bar c}, \,F}^{(0)} \right]
= \f{d{\hat \sigma}_{c{\bar c}, \,F}^{(0)}}{M^2 \;d{\bf\Omega}} 
\,(x_1p_1, x_2p_2;
{\bf\Omega}; \as(M^2) )
\;\;,
\eeq
where $x_1p_1^\mu$ ($x_2p_2^\mu$) is the momentum of the parton $c$ (${\bar c}$).
In Eq.~(\ref{qtycross}), we have included the contribution of both
the $q{\bar q}$ annihilation\footnote{In the case of $q{\bar q}$ annihilation,
the notation $c=q,{\bar q}$ is not completely precise, since the quark and the
antiquark can have either equal or different flavour. The same comment applies to
the factor $\left[ H^F C_1 C_2 \right]_{c{\bar c};a_1a_2}$.}
channel ($c=q,{\bar q}$) and the gluon fusion 
channel ($c=g$); one of these two contributing channels may be absent 
(i.e. $\left[ d\sigma_{c{\bar c}, \,F}^{(0)} \right]=0$ in that channel), 
depending on the
specific final state $F$.

The factor $S_c(M,b)$ in Eq.~(\ref{qtycross}) is universal (process independent):
it does not depend on the produced final-state system $F$ and on its kinematics.
It only depends on the partonic channel that produces the cross section
$\left[ d\sigma_{c{\bar c}, \,F}^{(0)} \right]$. Thus, $S_c(M,b)$ is called
quark ($c=q$ or ${\bar q}$) or gluon ($c=g$) Sudakov form factor in the cases of 
the $q{\bar q}$ annihilation or gluon fusion channel, respectively. 
The Sudakov form factor can be expressed in the following exponential form:
\begin{equation}
\label{formfact}
S_c(M,b) = \exp \left\{ - \int_{b_0^2/b^2}^{M^2} \frac{dq^2}{q^2} 
\left[ A_c(\as(q^2)) \;\ln \frac{M^2}{q^2} + B_c(\as(q^2)) \right] \right\} 
\;\;, 
\end{equation}
where the functions $A_c(\as)$ and $B_c(\as)$ are perturbative series in 
$\as$:
\beeq
\label{aexp}
A_c(\as) &=& \sum_{n=1}^\infty \left( \frac{\as}{\pi} \right)^n A_c^{(n)} 
\;\;, \\
\label{bexp}
B_c(\as) &= &\sum_{n=1}^\infty \left( \frac{\as}{\pi} \right)^n B_c^{(n)}
\;\;.
\eeeq

The factor $\left[ H^F C_1 C_2 \right]$ in Eq.~(\ref{qtycross})
has the following explicit form:
\beq
\label{what}
\left[ H^F C_1 C_2 \right]_{c{\bar c};a_1a_2}
 = H_c^F(x_1p_1, x_2p_2; {\bf\Omega}; \as(M^2))
\;\, C_{c \,a_1}(z_1;\as(b_0^2/b^2)) 
\;\, C_{{\bar c} \,a_2}(z_2;\as(b_0^2/b^2)) \;\;,
\eeq
where $H_c^F$ and $C_{a \,b}$ are both functions of $\as$, and they can be
perturbatively expanded as follows:
\beeq
\label{hexp}
H_c^F(x_1p_1, x_2p_2; {\bf\Omega}; \as) &= &1+ \sum_{n=1}^\infty 
\left( \frac{\as}{\pi} \right)^n 
H_c^{F \,(n)}(x_1p_1, x_2p_2; {\bf\Omega})
\;\;, \\
\label{cexp}
C_{a \,b}(z;\as) &=& \delta_{a \,b} \;\,\delta(1-z) + 
\sum_{n=1}^\infty \left( \frac{\as}{\pi} \right)^n C_{a \,b}^{(n)}(z) \;\;.
\eeeq
The function $H_c^{F}$ is process dependent, whereas
the perturbative functions $C_{a \,b}$ are universal 
and depend only on the parton indices $a$ and $b$ (analogously to the
dependence of the anomalous dimensions $\gamma_{a \,b}$ that control the
perturbative evolution of the parton densities, through the
Altarelli--Parisi equations).

By inspection of the right-hand side of Eq.~(\ref{what}), we notice 
that the scale of $\as$ is not set to a unique value. We have 
$\as(M^2)$ in the case of the function $H_c^{F}$ (as naturally expected for a
process-dependent contribution), and $\as(b_0^2/b^2)$ in the case of the 
functions $C_{c \,a_1}$ and $C_{{\bar c} \,a_2}$. The replacement 
$\as(M^2) \to \as(b_0^2/b^2)$ in Eq.~(\ref{what}) is feasible, provided it is
properly compensated [\ref{Catani:2000vq}] by a corresponding factor to be
inserted in Eq.~(\ref{formfact}): this procedure leads to a modification of the
Sudakov form factor, which becomes a process-dependent quantity.
Indeed, the hard-scattering function $H_c^{F}$ was introduced in 
Ref.~[\ref{Catani:2000vq}] to explicitly show 
(see also Ref.~[\ref{deFlorian:2000pr}]) the universality (process independence)
of both the Sudakov form factor and the coefficient function 
$C_{a\,b}$. In the version of Eq.~(\ref{qtycross}) that was originally
presented for the DY process (see Eq.~(1.1) in Ref.~[\ref{Collins:1984kg}]), 
the function $H_c^F=H_c^{DY}$ (note that $H_c^{DY}$ depends only on $\as$)
is absorbed in the definition of the functions $C_{a\,b}$ and of the function
$B_c$ of the form factor (these functions are thus `those' of the DY process).
As shown in Sect.~\ref{sec:ggfus}, the process-dependent 
hard-scattering function
$H_c^{F}$ definitely plays a distinctive role in the case of gluon fusion
subprocesses.

The present knowledge of the perturbative coefficients in Eqs.~(\ref{aexp}),
(\ref{bexp}), (\ref{hexp}) and (\ref{cexp}) is as follows
(see also Sect.~2.3 of Ref.~[\ref{Bozzi:2005wk}], where we used the same 
notation as in the present paper).
The coefficients $A_c^{(1)}$, $B_c^{(1)}$ and $A_c^{(2)}$ 
are known since a long time for both the quark [\ref{Kodaira:1981nh}] 
and the gluon [\ref{Catani:vd}] 
form factors. 
The explicit expression of the coefficient $B^{(2)}$ for the DY process
was first presented in Ref.~[\ref{Davies:1984sp}]. 
The process-independent structure and the explicit form of the 
coefficients $B_c^{(2)}$ ($c=q,g$) was derived 
in Ref.~[\ref{deFlorian:2000pr}].
The result of $A_c^{(3)}$ has been obtained very recently 
[\ref{Becher:2010tm}] by relating its value to the coefficient $B_c^{(2)}$
[\ref{Davies:1984sp}, \ref{deFlorian:2000pr}] and to the coefficient of 
the soft part of the Altarelli--Parisi
splitting functions at ${\cal O}(\as^3)$ [\ref{Vogt:2004mw}].
The universal first-order coefficients $C_{q \,g}^{(1)}(z)$ and 
$C_{g \,q}^{(1)}(z)$ were first computed in 
Refs.~[\ref{Davies:1984hs}, \ref{Collins:1984kg}] and 
Ref.~[\ref{Kauffman:1991cx}], respectively.
The general result for the first-order coefficients
$H_c^{F \,(1)}$ and $C_{a \,b}^{(1)}(z)$ was derived in 
Ref.~[\ref{deFlorian:2000pr}], where the process dependence of 
$H_c^{F \,(1)}$ is explicitly related to the first-order virtual corrections
of the partonic subprocesses $c + {\bar c} \to F$ in Eqs.~(\ref{ann}) 
and (\ref{fus}). The coefficients of Eq.~(\ref{what}) at the second order in
$\as$ have been computed for both SM Higgs boson production by gluon
fusion\footnote{See, however, 
related comments in Sect.~\ref{sec:ggfus}.} [\ref{Catani:2007vq}] 
and the DY process [\ref{Catani:2009sm}, \ref{Bozzi:2010xn}].

We note that the resummation formula (\ref{qtycross}) involves perturbative
functions of $\as(\mu^2)$, where the scale $\mu^2$ is 
$\mu^2=M^2$ (see Eqs.~(\ref{sig0}) and (\ref{what})), or $\mu^2=q^2$
(see Eq.~(\ref{formfact})), or $\mu^2=b_0^2/b^2$ (see Eq.~(\ref{what})).
All these functions can be expressed in terms of $\as(\mu_R^2)$, 
$\ln (\mu^2/\mu_R^2)$ and the 
perturbative coefficients of the QCD $\beta$-function 
($\mu_R$ is the {\em renormalization scale} that customarily appears in
fixed-order calculations) by using the renormalization group equation for the
perturbative evolution of the QCD running coupling  $\as(\mu^2)$.

We add a relevant (though known) observation. Considering the dependence on the
impact parameter $\bf b$, all the factors in the integrand of the Fourier
transformation on the right-hand side of the resummation formula
(\ref{qtycross}) are functions of ${\bf b}^2$, with no 
dependence on the azimuthal
angle $\phi({\bf b})$ 
of $\bf b$ in the transverse plane of the collision.
Therefore, in Eq.~(\ref{qtycross}) we can straightforwardly perform the
integration over $\phi({\bf b})$ and implement the replacement
\beq
\label{fourbes}
\int \f{d^2{\bf b}}{2\pi} \;\, e^{i {\bf b}\cdot \bqt} \;\;F({\bf b}^2)
= \int_0^{+\infty} \,db \;b  \;J_0(b q_T) \;\;F(b^2) \;\;,
\eeq
where $J_0(x)$ is the $0$th-order Bessel function, and $F({\bf b}^2)$
denotes a generic function of ${\bf b}^2$.
This result implies a technical simplification of the resummation formula,
since the {\em two}-dimensional Fourier transformation is replaced by the
{\em one}-dimensional Bessel transformation. More importantly, this implies that
the right-hand side of Eq.~(\ref{qtycross}) depends only on $\bqt^{\!2}$,
with no additional dependence on the 
azimuthal angle $\phi(\bqt)$ 
of $\bqt$. Therefore, according to Eqs.~(\ref{qtycross}) and
(\ref{fourbes}), the singular part of the
$\bqt$ differential cross section in Eq.~(\ref{singsig}) does not contain {\em
any} azimuthal correlations with respect to $\bqt$. Equivalently, we can say 
that $d\sigma_F/d^2\bqt$ and $d\sigma_F/d\qt^2$ are simply proportional,
and we can write:
\beq
\label{singsigqbarq}
\f{d\sigma_F^{({\rm sing})}}{d^2{\bqt} 
\;dM^2 \;dy \;d{\bf\Omega}} 
= \f{1}{\pi} \;\f{d\sigma_F^{({\rm sing})}}{d\qt^2 
\;dM^2 \;dy \;d{\bf\Omega}} 
\;\;.
\eeq
Obviously, this does not mean that the multidifferential cross section 
$d\sigma_F/d^2\bqt$ in Eq.~(\ref{diffxs}) has no azimuthal correlations. 
In general,
azimuthal correlations are present in the regular part $d\sigma_F^{({\rm reg})}$
(see Eq.~(\ref{Fdec})) of the cross section. In the small-$\qt$ region ($\qt \ll
M$), $d\sigma_F^{({\rm reg})}$ is of ${\cal O}(\qt/M)$ (modulo powers of 
$\ln(M^2/\qt^2)$)
with respect to 
$d\sigma_F^{({\rm sing})}$ order-by-order in QCD perturbation theory. In the case
of gluon fusion processes, these conclusions about azimuthal correlations at small
$\qt$ are no
longer true in view of the results presented in the next section.

\section{Transverse-momentum resummation
in gluon fusion\\
 processes}
\label{sec:ggfus}

The resummation formalism reviewed in Sect.~\ref{sec:revqqbar}
was originally developed and proven 
[\ref{Dokshitzer:hw}--\ref{Collins:1984kg}]
for the DY process (and related observables, such as the energy--energy 
correlation function in $e^+e^-$ annihilation). 
In the subsequent literature these 
results were extrapolated to various processes of the class 
in Eq.~(\ref{process}). In this section, we show that this '\naive'
extrapolation is not valid in the case of the (sub)processes that are 
controlled by gluon fusion.

To present our results, we start from the resummation formula in 
Eq.~(\ref{qtycross}). It includes the contributions from both $q{\bar q}$
annihilation $(c=q,{\bar q})$ and gluon fusion $(c=g)$. We thus separate these
two types of contributions, and we write:
\beq
\label{qgdec}
\left[ d\sigma_F \right] = \left[ d\sigma_F \right]^{(q{\bar q}-{\rm ann.})}
+ \left[ d\sigma_F \right]^{(g-{\rm fus.})}
\;\;.
\eeq
Our new results refer to $\left[ d\sigma_F \right]^{(g-{\rm fus.})}$, whereas
we maintain the results of Sect.~\ref{sec:revqqbar} for 
$\left[ d\sigma_F \right]^{(q{\bar q}-{\rm ann.})}$. To be precise, also in the
case of gluon fusion, we confirm the factorization structure on the right-hand
side of Eq.~(\ref{qtycross}). We explicitly report this structure:
\beeq
\label{qtycrossgg}
&&\left[ d\sigma_F \right]^{(g-{\rm fus.})} =\f{M^2}{s} \;
 \left[ d\sigma_{gg, \,F}^{(0)} \right]
\int \f{d^2{\bf b}}{(2\pi)^2} \;\, e^{i {\bf b}\cdot \bqt} \;
  S_g(M,b)\nn \\
&& \;\;\;\; \times \;
\sum_{a_1,a_2} \,
\int_{x_1}^1 \f{dz_1}{z_1} \,\int_{x_2}^1 \f{dz_2}{z_2} 
\; \left[ H^F C_1 C_2 \right]_{gg;a_1a_2}
\;f_{a_1/h_1}(x_1/z_1,b_0^2/b^2)
\;f_{a_2/h_2}(x_2/z_2,b_0^2/b^2) \;
\;, 
\eeeq
where the lowest-order cross section $\left[ d\sigma_{gg, \,F}^{(0)} \right]$
and the gluon form factor $S_g(M,b)$ are given in Eqs.~(\ref{sig0}) and 
(\ref{formfact}), respectively. The new results regard the gluon fusion factor 
$\left[ H^F C_1 C_2 \right]_{gg;a_1a_2}$.

The \naive\  expression on the right-hand side of Eq.~(\ref{what}) has to be
replaced by the following result [\ref{CGinprep}]:
\beeq
\label{whatgg}
\left[ H^F C_1 C_2 \right]_{gg;a_1a_2}
&=& H_{\mu_1 \,\nu_1, \mu_2 \,\nu_2 }^F(x_1p_1, x_2p_2; {\bf\Omega}; \as(M^2))
\nn \\
&\times&\; C_{g \,a_1}^{\mu_1 \,\nu_1}(z_1;p_1,p_2,{\bf b};\as(b_0^2/b^2)) 
\;\, C_{g \,a_2}^{\mu_2 \,\nu_2}(z_2;p_1,p_2,{\bf b};\as(b_0^2/b^2)) 
\;\;. 
\eeeq
The first evident difference with respect to Eq.~(\ref{what}) is the presence of
Lorentz tensors, rather than scalar functions. The Lorentz indices (symbolically)
refer to the gluon fusion hard-scattering process
\beq
\label{gglor}
g_{(\mu_1)}(x_1 p_1) + g_{(\mu_2)}(x_2 p_2) 
\to F \;\;,
\eeq
where $\mu_i$ $(i=1,2)$ is the Lorentz index carried by the external gluon leg
with incoming momentum $x_i p_i$. The indices $\nu_1$ and $\nu_2$ refer to the 
external gluon legs of the process that is complex conjugate to that in
Eq.~(\ref{gglor}).

The process-dependent factor $H^F$ in Eq.~(\ref{whatgg}) has the following
perturbative expansion:
\beeq
\label{hexpgg}
H_{\mu_1 \,\nu_1, \mu_2 \,\nu_2 }^F(x_1p_1, x_2p_2; {\bf\Omega}; \as) 
&=& H_{\mu_1 \,\nu_1, \mu_2 \,\nu_2 }^{F \,(0)}(x_1p_1, x_2p_2; {\bf\Omega})
\nn \\
&+& \sum_{n=1}^\infty 
\left( \frac{\as}{\pi} \right)^n 
H_{\mu_1 \,\nu_1, \mu_2 \,\nu_2 }^{F \,(n)}(x_1p_1, x_2p_2; {\bf\Omega})
\;\;, 
\eeeq
with the lowest-order constraint:
\beq
\label{h0gg}
H_{\mu_1 \,\nu_1, \mu_2 \,\nu_2 }^{F \,(0)} \;g^{\mu_1 \,\nu_1} 
\;g^{\mu_2 \,\nu_2} = 1 \;\;.
\eeq
We also define the scalar function $H_g^F$ as follows:
\beq
\label{hgg}
H_g^F(x_1p_1, x_2p_2; {\bf\Omega}; \as) \equiv
H_{\mu_1 \,\nu_1, \mu_2 \,\nu_2 }^{F}(x_1p_1, x_2p_2; {\bf\Omega}; \as) 
\;g^{\mu_1 \,\nu_1}  \;g^{\mu_2 \,\nu_2}  \;\;.
\eeq
A relevant property [\ref{CGinprep}] of the Lorentz tensor
$H_{\mu_1 \,\nu_1, \mu_2 \,\nu_2 }^{F}$ is current conservation, namely:
\beeq
\label{curcons}
&&p_1^{\mu_1} 
\;H_{\mu_1 \,\nu_1, \mu_2 \,\nu_2 }^{F}(x_1p_1, x_2p_2; {\bf\Omega}; \as)
= p_1^{\nu_1} 
\;H_{\mu_1 \,\nu_1, \mu_2 \,\nu_2 }^{F}(x_1p_1, x_2p_2; {\bf\Omega}; \as)
= 0 \;\;, \nn \\
&&p_2^{\mu_2} 
\;H_{\mu_1 \,\nu_1, \mu_2 \,\nu_2 }^{F}(x_1p_1, x_2p_2; {\bf\Omega}; \as)
= p_2^{\nu_2} 
\;H_{\mu_1 \,\nu_1, \mu_2 \,\nu_2 }^{F}(x_1p_1, x_2p_2; {\bf\Omega}; \as)
= 0 \;\;.
\eeeq

The universal (process-independent) partonic tensor 
$C^{\mu_i \,\nu_i}$ in Eq.~(\ref{whatgg}) exhibits an explicit dependence
on the impact parameter ${\bf b}$, besides the implicit dependence on $b^2$
through the scale of $\as$. The structure of the partonic tensor is:
\beq
\label{cggten}
C_{g \,a}^{\,\mu \nu}(z;p_1,p_2,{\bf b};\as) =
d^{\,\mu \nu}(p_1,p_2) \;C_{g \,a}(z;\as) + D^{\,\mu \,\nu}(p_1,p_2;{\bf b}) 
\;G_{g \,a}(z;\as) \;\;,
\eeq
where
\beq
\label{dten}
d^{\,\mu \nu}(p_1,p_2) = - \,g^{\mu \nu} + 
\f{p_1^\mu p_2^\nu+ p_2^\mu p_1^\nu}{p_1 \cdot p_2} \;\;,
\eeq
\beq
\label{dbten}
D^{\,\mu \nu}(p_1,p_2;{\bf b}) = d^{\,\mu \nu}(p_1,p_2) - 
2 \; \f{b^\mu \,b^\nu}{\bf b^2} \;\;,
\eeq
and $b^\mu = (0,{\bf b},0)$ is the two-dimensional impact parameter vector
in the four-dimensional notation $(b^\mu b_\mu = - {\bf b^2})$.
The gluonic coefficient function $C_{g \,a}(z;\as)$ 
has the same perturbative structure as in Eq.~(\ref{cexp}).
The first-order coefficient $C_{g \,a}^{(1)}(z)$ of the function 
$C_{g \,a}(z;\as)$ in Eq.~(\ref{cggten}) and the 
first-order coefficient $H_g^{F \,(1)}$ of the function $H_g^F$ in 
Eq.~(\ref{hgg}) have actually the same value [\ref{deFlorian:2000pr}]
as obtained in the context of the `\naive' expression in Eq.~(\ref{what}).

The partonic coefficient function $G_{g \,a}(z;\as)$ in Eq.~(\ref{cggten})
is a specific and distinctive feature of transverse-momentum resummation in gluon
fusion processes\footnote{Setting $G_{g \,a}(z;\as)=0$ in Eq.~(\ref{cggten}),
the gluon fusion factor of Eq.~(\ref{whatgg}) coincides with the corresponding
`\naive' factor of Eq.~(\ref{what}).}. 
Its perturbative expansion starts at order $\as$, and we write: 
\beq
\label{gfexp}
G_{g \,a}(z;\as) = \f{\as}{\pi} \;G_{g \,a}^{(1)}(z) \,+
\sum_{n=2}^\infty \left( \frac{\as}{\pi} \right)^n G_{g \,a}^{(n)}(z) \;\;.
\eeq
The first-order coefficients are [\ref{CGinprep}]
\beeq
\label{gg1}
&&G_{g \,g}^{(1)}(z) = C_A \;\f{1-z}{z} \;\;, \\
\label{gq1}
&&G_{g \,q}^{(1)}(z) = G_{g \,{\bar q}}^{(1)}(z) = C_F \;\f{1-z}{z} \;\;.
\eeeq
The second-order coefficients $C_{g \,a}^{(2)}(z)$ and 
$G_{g \,a}^{(2)}(z)$ are considered in 
Refs.~[\ref{Catani:2007vq}, \ref{CGinprep}].

The tensors in Eqs.~(\ref{dten}) and (\ref{dbten}) fulfil the relations
$g^{\mu \nu} d_{\mu \nu}= -2$ and $g^{\mu \nu} D_{\mu \nu}= 0$.
Considering the centre--of--mass system of the collision and denoting by
$\mu=1,2$ the Lorentz indices of the non-vanishing components of
purely-transverse vectors (such as $\bqt$ and $\bf b$), we have 
$d_{\mu \nu}= D_{\mu \nu}= 0$ if $\mu=0 \,(\nu=0)$ or $\mu=3 \,(\nu=3)$.
Therefore, the only non-vanishing components of 
$d_{\mu \nu}$ and $D_{\mu \nu}$ are those that correspond to Lorentz indices
$j=1,2$ and $k=1,2$ of the transverse plane; we have
\beq
\label{dtran}
d^{\,j k}(p_1,p_2) = - \,g^{\,j k} \;\;, \quad
D^{\,j k}(p_1,p_2;{\bf b}) = - \,g^{\,j k} - 
2 \; \f{b^j \,b^k}{\bf b^2} \;\;.
\eeq

The structure of Eqs.~({\ref{whatgg}) and ({\ref{cggten})
has a definite physical origin: this structure is produced by 
{\em gluon collinear correlations} [\ref{CGinprep}]. We refer to the
correlations that occur in the universal (process-independent) partonic 
subprocess
\beq
\label{gcoll}
a \;\to \;g + a_1 +a_2 + \dots \;\;, 
\eeq 
where the initial-state colliding parton $a$ ($a=q,{\bar q},g$) `evolves'
in the colliding gluon $g$ through {\em collinear} radiation of the final-state
partons $a_1, a_2, \dots\;\,$.

To illustrate the role of gluon collinear correlations, we briefly sketch how
they arise at the first non-trivial order in QCD perturbation theory.
We consider the partonic hard-scattering process
\beq
\label{radpro}
a(p) + g({\bar p}) \;\to \; F(q) + a(k) \;\;,
\eeq 
which leads to the first-order (real) radiative corrections to the gluon fusion
process in Eq.~(\ref{fus}), namely,
\beq
\label{progg}
g(p) + g({\bar p}) \;\to \; F(q)  \;\;.
\eeq 
Here the parton momenta are denoted by $p, {\bar p}$ and $k$.
In the process (\ref{progg}), the final-state system $F$ is produced with a
vanishing transverse momentum. The final-state system $F$ acquires a 
non-vanishing transverse momentum $\qt^\mu$ $(\qt^\mu \,q_{T \mu} = - \bqt^2)$
through the radiative process in Eq.~(\ref{radpro}); here, the final-state parton
$a$ has transverse momentum $k_T^\mu$, and momentum conservation implies
$k_T^\mu = -\qt^\mu$. In the small-$\qt$ region (formally, when $\qt \to 0$),
the scattering amplitude of the process in Eq.~(\ref{radpro}) is singular, and
the singular behaviour is controlled by a well-known universal factorization
formula (see, e.g., Eq.~(4.23) and related formulae in Sect.~4.3 of 
Ref.~[\ref{Catani:1996vz}]).
The factorization formula relates the scattering amplitude of the process 
in Eq.~(\ref{radpro}) with the scattering amplitude of the 
hard-scattering subprocess $g(zp) + g({\bar p}) \to F$.
The singular factor is produced by the 
transverse-momentum spectrum of the collinear splitting subprocess
\beq
\label{gcoll1}
a(p) \;\to \;g(zp) + a((1-z)p) \;\;, 
\eeq 
where $z$ is the longitudinal-momentum fraction that is transferred from the
initial-state parton $a$ to the initial-state gluon $g$.
At the lowest perturbative order in the QCD coupling, the  
transverse momentum spectrum of the subprocess in Eq.~(\ref{gcoll1}) is
proportional to
\beq
\label{spect}
\f{\as}{\pi} \;\f{d^2\bqt}{\bqt^2} \;dz 
\;\left[ {\hat P}^{(1)}_{g \,a}(z,\bqt) \right]^{\mu \,\nu} \;\;,
\eeq
where $k_T^\mu = -\qt^\mu$ is the transverse momentum of the the final-state
parton $a$ in Eq.~(\ref{gcoll1}), and the collinear splitting function 
$\left[ {\hat P}^{(1)}_{g \,a}(z,\bqt) \right]^{\mu \,\nu}$ has the following
explicit from:
\beq
\label{ggpol}
\left[ {\hat P}^{(1)}_{gg}(z,\bqt) \right]^{\mu \,\nu} = 2 \,C_A \left[
- \,g^{\,\mu \,\nu} \left( \f{z}{1-z} + z(1-z) \right) 
+ 2 \;\f{\qt^\mu \qt^\nu}{\bqt^{2}} \;\f{1-z}{z} \;\right] \;\;,
\eeq
 \beq
\label{gqpol}
\left[ {\hat P}^{(1)}_{gq}(z,\bqt) \right]^{\mu \,\nu} = 
\left[ {\hat P}^{(1)}_{g{\bar q}}(z,\bqt) \right]^{\mu \,\nu} =
2 \,C_F \left[
- g^{\,\mu \,\nu} \;\f{1}{2} \;z + 2 \;\f{\qt^\mu \qt^\nu}{\bqt^{2}} \;\f{1-z}{z}
\;\right] \;\;.
\eeq 
In Eq.~(\ref{spect}),
$\mu$ and $\nu$ are the Lorentz indices of the gluon $g(zp)$ in the process 
(\ref{gcoll1}) and in its complex conjugate process, respectively.

In the context of our present study, the most important feature of the gluonic
splitting process in Eq.~(\ref{gcoll1}) is that it is 
intrinsically {\em polarized}.
The corresponding collinear splitting function 
$\left[ {\hat P}^{(1)}_{g \,a}(z,\bqt) \right]^{\mu \,\nu}$ has a non-trivial
dependence on the Lorentz (and, thus, spin) indices of the gluon, and this
dependence is controlled by the azimuthal angle of the small transverse 
momentum that is radiated in the splitting process. We remark that this
polarization effect
is present despite the fact that we are not considering polarized-scattering
processes. Indeed, we have performed the {\em sum} over the spin polarizations of
the final-state parton $a((1-z)p)$ {\em and} the {\em average} over the spin
polarizations of the initial-state parton $a(p)$. The intrinsic gluon
polarization effects that arise in the splitting process (\ref{gcoll1}) (and, 
more generally, in the collinear splitting process of Eq.~(\ref{gcoll}))
produce {\em correlations} between 
the initial-state gluon legs of the scattering amplitude of the
factorized hard-scattering subprocess
$g(zp) + g({\bar p}) \to F$ and of the corresponding complex conjugate 
scattering amplitude.

Replacing the gluon $g$ with a quark $q$ (or an antiquark ${\bar q}$)
in Eq.~(\ref{gcoll}), we obtain the quark (antiquark) collinear evolution 
process
\beq
\label{qcoll}
a \;\to \;q({\bar q}) + a_1 +a_2 + \dots \;\;. 
\eeq 
In this case, if we sum over the spin polarizations of the final-state partons
$a_i$ and we average over the spin polarizations of the initial-state parton $a$,
the collinear evolution of the quark (antiquark) turns out to be {\em
unpolarized}, as a consequence of helicity conservation in QCD radiation from
a massless quark (antiquark). This essential difference between the collinear
evolution of quarks (antiquarks) and gluons is eventually the origin of the
difference of transverse-momentum resummation between $q{\bar q}$ annihilation
processes and gluon fusion processes.

Going back to Eqs.~(\ref{spect})--(\ref{gqpol}), we can rewrite the gluon
splitting functions as follows:
\beq
\label{gapol}
\left[ {\hat P}^{(1)}_{g \,a}(z,\bqt) \right]^{\mu \,\nu} = - \,g^{\,\mu \,\nu}
\;{\hat P}^{(1)}_{g \,a}(z) + 
\left( g^{\,\mu \,\nu} + 2 \;\f{\qt^\mu \qt^\nu}{\bqt^{2}}
\right)
2 \;G^{(1)}_{g \,a}(z) \;\;, \quad a=g,q,{\bar q} \;\;,
\eeq
where the functions $G^{(1)}_{g \,a}(z)$ are those in Eqs.~(\ref{gg1}) $(a=g)$
and (\ref{gq1}) $(a=q,{\bar q})$, and the functions 
${\hat P}^{(1)}_{g \,a}(z)$ are
\beq
\label{pgg}
{\hat P}^{(1)}_{g \,g}(z) = 2 \,C_A \left[ \f{z}{1-z} + \f{1-z}{z} + z(1-z)
\right] \;\;,
\eeq
\beq
\label{pgq}
{\hat P}^{(1)}_{g \,q}(z) = {\hat P}^{(1)}_{g \,{\bar q}}(z) 
= C_F \;\f{1 + (1-z)^2}{z} \;\;.
\eeq
We see that ${\hat P}^{(1)}_{g \,a}(z)$ are the (real part of) first-order
Altarelli--Parisi probabilities that control the customary collinear evolution
of the unpolarized gluon parton density $f_{g/h}(x,\mu^2)$.
In Eq.~(\ref{gapol}), the unpolarized splitting functions 
${\hat P}^{(1)}_{g \,a}(z)$ are multiplied by the tensor\footnote{This tensor can
be replaced by the tensor $d^{\,\mu \nu}(p,{\bar p})$ of Eq.~(\ref{dten}),
because of current conservation (gauge invariance) of the scattering amplitude
of the factorized hard-scattering subprocess
$g(zp) + g({\bar p}) \to F$.}
$-\,g^{\,\mu \,\nu}$
that does not produce any gluonic correlations; the collinear correlation
effects are produced by the tensor that multiplies the functions  
$G^{(1)}_{g \,a}(z)$ in Eq.~(\ref{gapol}). Using Eq.~(\ref{gapol}) and 
performing the Fourier transformation
of Eq.~(\ref{spect}) from $\bqt$ space to $\bf b$ space, we reproduce the
lowest-order structure of the resummation formulae in Eqs.~(\ref{qtycrossgg}) 
and (\ref{whatgg}) (more details are given in Ref.~[\ref{CGinprep}]).
The term proportional to the unpolarized splitting 
function ${\hat P}^{(1)}_{g \,a}(z)$ produces the
evolution of the parton density $f_{g/h}(x,\mu^2)$ up to the scale
$\mu^2=b_0^2/b^2$; this term also produces [\ref{deFlorian:2000pr}]
a residual effect (included in the
function $C_{g \,a}(z;\as)$ of Eq.~(\ref{cggten})) that is related to the
definition of the parton densities in the \ms\  factorization scheme.
The term proportional to $G^{(1)}_{g \,a}(z)$ produces the first-order
contribution to the function $G_{g \,a}(z;\as)$ in Eq.~(\ref{cggten}).

The factorized structure of Eq.~(\ref{whatgg}) has no direct interpretation in
terms of a probabilistic (classical or quasi-classical) partonic picture.
In fact, the structure originates at the cross section level from an underlying
quantum-mechanical {\em interference} between scattering amplitudes and their
complex conjugates. In particular, the gluonic tensor 
$C_{g \,a}^{\,\mu \nu}$ in Eqs.~(\ref{whatgg}) and (\ref{cggten})
(unlike the quark coefficient $C_{q \,a}$ in Eq.~(\ref{what})) cannot be
interpreted as the simple residual 
(and factorization-scheme dependent)
effect of the customary evolution of the parton density 
$f_{a/h}(x,\mu^2)$ (which is a Lorentz scalar) from a non-perturbative scale up to
the scale $\mu^2=b_0^2/b^2$.
Incidentally, we note that the first-order functions $G_{g \,a}^{(1)}(z)$
$(a=g,q,{\bar q})$ in Eqs.~(\ref{gg1}) and (\ref{gq1})
do not depend on the factorization scheme of the parton densities.

The interference phenomenon that leads to Eq.~(\ref{whatgg}) produces specific
physical effects. The tensor $D_{\mu \nu}$ (see Eqs.~(\ref{cggten}) and 
(\ref{dbten})) explicitly depends on the direction of the impact parameter 
vector $\bf b$ in the transverse plane. This dependence produces enhanced
(i.e. non-suppressed by terms of ${\cal O}(\qt/M)$) {\em spin} and {\em
azimuthal} correlations in the $\bqt$ differential cross section $d\sigma^F$
at small values of $\qt$. The spins of the gluons in the gluon fusion process of
Eq.~(\ref{gglor}) are correlated through Eq.~(\ref{whatgg}). The azimuthal 
angles of the particles of the final-state system $F$ are correlated to the 
azimuthal angle of $\bqt$ through Eq.~(\ref{whatgg}) and the Fourier
transformation in Eq.~(\ref{qtycrossgg}).

Note that spin and azimuthal correlations do not necessarily show up their
effect simultaneously. For instance, considering the differential cross section 
averaged over the azimuthal angle of $\bqt$, the azimuthal correlations
cancel, whereas the spin correlations can survive. This observation also
implies that, in gluon fusion processes, the \naive\ resummation factor of 
Eq.~(\ref{what}) is not correct even in the case of the azimuthally-averaged
cross section $d\sigma_F/d\qt^2$ (see Sect.~\ref{sec:azav}). 
To explicitly illustrate
this observation in a very simple manner, below we consider a specific process:
the production of the SM Higgs boson, $H$, by the gluon fusion mechanism.
Since $H$ is a scalar particle, its $\bqt$ cross section has no 
azimuthal correlations.

In the case of Higgs boson production, the final-state system $F$ in 
Eq.~(\ref{process}) is simply $H$, with momentum $q^\mu$. Within the SM,
the corresponding gluon fusion production mechanism is mediated by a heavy-quark
(mainly, top quark) loop. Our conclusions are unchanged if we consider a generic
$ggH$ effective coupling (such as, for instance, the SM effective coupling
that is obtained in the large-$M_t$ approximation, $M_t$ being the mass of the
top quark). Since $H$ is a scalar particle of spin~$0$, the tensor structure of
the corresponding hard factor 
$H_{\mu_1 \,\nu_1, \mu_2 \,\nu_2 }^{F=H}$ in Eq.~(\ref{whatgg}) is uniquely
determined by Lorentz covariance, parity conservation and gauge invariance;
we have
\beq
\label{hh}
H_{\mu_1 \,\nu_1, \mu_2 \,\nu_2 }^{F=H}(x_1p_1,x_2p_2;\as(M^2))
= \f{1}{2} \;H_{g}^{F=H}(\as(M^2)) 
\;\left( g_{\mu_1 \mu_2} - \f{p_{2 \,\mu_1} p_{1 \,\mu_2}}{p_1 \cdot p_2}\right)
\;\left( g_{\nu_1 \nu_2} - \f{p_{2 \,\nu_1} p_{1 \,\nu_2}}{p_1 \cdot p_2}\right)
\;\;,
\eeq
where the scalar function $H_{g}^{F=H}$ (see Eq.~(\ref{hgg}))
only depends on $\as$ (apart from the dependence on $M_t/M$ or other parameters
of the effective coupling $ggH$).
Using Eqs.~(\ref{cggten}) and (\ref{hh}),
the Higgs boson resummation factor of Eq.~(\ref{whatgg}) is
\beeq
\label{whath}
\left[ H^{F=H} C_1 C_2 \right]_{gg;a_1a_2}
&=& H_{g}^{F=H}(\as(M^2)) \;\left[ \; C_{g \,a_1}(z_1;\as(b_0^2/b^2)) 
\;\, C_{g \,a_2}(z_2;\as(b_0^2/b^2)) \right.
\nn \\
&+& \left.  G_{g \,a_1}(z_1;\as(b_0^2/b^2)) 
\;\, G_{g \,a_2}(z_2;\as(b_0^2/b^2)) \;
\right]
\;\;. 
\eeeq
Note that the right-hand side does not depend on the direction of $\bf b$:
this implies that the $\bqt$ distribution has no azimuthal correlations in the
small-$\qt$ region. This result is consistent with the fact that the 
$\bqt$ cross section for $H$ production has no azimuthal correlations
at any values of $\qt$.

We can compare the result in Eq.~(\ref{whath}) with the structure in 
Eq.~(\ref{what}). Owing to the specific factorized dependence on 
$z_1$ and $z_2$ of the two terms in the square bracket of Eq.~(\ref{whath}),
the term proportional to $G(z_1) \,G(z_2)$ cannot be removed by a redefinition 
of the function $C(z;\as)$. This shows that the \naive\ gluon fusion expression
in Eq.~(\ref{what}) is not valid even in the simple case of Higgs boson
production. The structure of Eq.~(\ref{whath}) obviously differs from the
(correct) structure of Eq.~(\ref{what}) for $q{\bar q}$ annihilation processes.
The additional term proportional to $G(z_1) \,G(z_2)$ is indeed produced by
(gluon) spin correlations (see also Sect.~\ref{sec:hel}), which have no analogue
in $q{\bar q}$ annihilation processes. We finally observe that the term
$G(z_1) \,G(z_2)$ in Eq.~(\ref{whath}) starts at ${\cal O}(\as^2)$ in QCD
perturbation theory (see Eq.~(\ref{gfexp})). Therefore, its effect first appears
in the computation of the next-to-next-to-leading order (NNLO) QCD radiative
corrections to the Higgs boson $\qt$ cross section.

\setcounter{footnote}{1}

We now comment on our paper in Ref.~[\ref{Catani:2007vq}]. The paper deals with
the class of processes in Eq.~(\ref{process}) and presents a practical formalism
for the NNLO QCD calculation of the corresponding cross sections at the
fully-differential level. The formalism exploits the subtraction method to cancel
the unphysical infrared (IR) divergences 
that separately occur in the real and
virtual radiative corrections. The explicit construction of the  
subtraction counterterms is based on the universal structure of
transverse-momentum resummation formulae and on their expansion up to NNLO in 
QCD perturbation theory. These are the resummation formulae discussed in the
present paper. Working on the research project of 
Ref.~[\ref{Catani:2007vq}], we found the results presented in this section,
and explicitly documented for the first time in this paper. These results are
essential for the application of the subtraction method of 
Ref.~[\ref{Catani:2007vq}] to gluon fusion processes. In these processes,
the \naive\ expression in Eq.~(\ref{what}) does not reproduce the correct (and
singular) perturbative behaviour of the $\qt$ cross section in the limit 
$\qt \to 0$. Using Eq.~(\ref{what}), rather than the correct result in  
Eq.~(\ref{whatgg}), leads to subtraction counterterms that would spoil the 
cancellation of the IR divergences. Although the theoretical results of the
present paper were not explicitly illustrated in Ref.~[\ref{Catani:2007vq}],
they were actually taken into account in the NNLO computations presented
therein. In particular, the explicit application to Higgs boson production
(which was implemented in the Monte Carlo code HNNLO) considered in 
Ref.~[\ref{Catani:2007vq}] is based on and implements the results (e.g.
Eq.~(\ref{whath})) illustrated in this section.

We also add a brief comment on Ref.~[\ref{Mantry:2009qz}]. The authors of  
Ref.~[\ref{Mantry:2009qz}] study transverse-momentum cross sections at small 
$\qt$ by introducing a factorization formalism that differs from the 
resummation formalism considered in Sect.~\ref{sec:revqqbar} 
and in this section. In the case of SM Higgs boson production or, more
generally, gluon fusion processes, a perturbative ingredient of the 
factorization
formulae in Refs.~[\ref{Mantry:2009qz}, \ref{Mantry:2010mk}]
is a collinear function ${\cal I}^{\mu \,\nu}_{g \,a}$
(the Lorentz indices $\mu$ and $\nu$ and the parton indices $g$ and $a$
refer to the notation used throughout our paper), which emerges from authors'
analysis based on Soft Collinear Effective Theory. The 
function ${\cal I}^{\mu \,\nu}_{g \,a}$ is explicitly computed 
[\ref{Mantry:2009qz}, \ref{Mantry:2010mk}] up to its first-order contribution in
$\as$. The result at ${\cal O}(\as)$ is expressed in terms of two form factors,
${\cal F}_1^{g \,a}$ and ${\cal F}_2^{g \,a}$, that multiply the tensors
in Eqs.~(\ref{dten}) and (\ref{dbten}), respectively.
Moreover, at order $\as$, we note 
(see\footnote{We assume that there is a typo in the overall sign
of the argument of the hypergeometric function ${}_0F_1$ in Eq.~(37) 
of Ref.~[\ref{Mantry:2010mk}].}
Eqs.~(35) and (37) in Ref.~[\ref{Mantry:2010mk}])
that the ratio 
${\cal F}_2^{g \,g}/{\cal F}_2^{g \,q}$ is equal to $C_A/C_F$;
the same value, $C_A/C_F$, is obtained by considering the ratio of 
our functions $G_{g \,g}^{(1)}(z)$
and $G_{g \,q}^{(1)}(z)$ in Eqs.~(\ref{gg1}) and (\ref{gq1}).
In Ref.~[\ref{Mantry:2009qz}], the authors also note that the form factor 
${\cal F}_2^{g \,a}$ does not contribute to the Higgs boson $\qt$ cross section
at the next-to-leading order (NLO). These first-order features of the 
function ${\cal I}^{\mu \,\nu}_{g \,a}$ have clear analogies with the structure
of our universal coefficient tensor $C_{g \,a}^{\,\mu \nu}$ 
in Eq.~(\ref{cggten}).

\section{
The gluon fusion resummation formula in helicity space}
\label{sec:hel}

The factorization formula in Eq.~(\ref{whatgg}) involves sums over the 
Lorentz indices of the gluons (i.e., of the gluon field $A_\mu$).
These sums can be replaced by corresponding sums over the spin polarization 
states
of the gluon. In particular, it can be convenient to consider physical
polarization states of
definite helicity $\lambda$ ($\lambda=\pm$). In this section we present the
helicity space version of the factorization formula
in Eq.~(\ref{whatgg}).

Exploiting gauge invariance or, more precisely, current conservation
(e.g. $p_{1 \, \mu_1} C_{g \,a_1}^{\mu_1 \,\nu_1}(z_1;\dots)$ and 
Eq.~(\ref{curcons})), the right-hand side of Eq.~(\ref{whatgg})
can be written in the following form:
\beeq
\label{whathel}
\left[ H^F C_1 C_2 \right]_{gg;a_1a_2}
&=& \sum_{\lambda_1, h_1,\lambda_2, h_2}
H_{(\lambda_1, h_1), (\lambda_2, h_2)}^F(x_1p_1, x_2p_2; {\bf\Omega}; \as(M^2))
\nn \\
&\times&\; C_{g \,a_1}^{(\lambda_1, h_1)}(z_1;p_1,p_2,{\bf b};\as(b_0^2/b^2)) 
\;\, C_{g \,a_2}^{(\lambda_2, h_2)}(z_2;p_1,p_2,{\bf b};\as(b_0^2/b^2)) 
\;\;, 
\eeeq
where $\lambda_i$ and $h_i$ are helicity space indices ($\lambda_i=\pm$,
$h_i=\pm$). The relation between the `helicity tensors' in Eq.~(\ref{whathel})
and the Lorentz tensors in Eq.~(\ref{whatgg}) is:
\beeq
\label{hhel}
H_{(\lambda_1, h_1), (\lambda_2, h_2)}^F(x_1p_1, x_2p_2; {\bf\Omega}; \as)
&\!\!=& H_{\mu_1 \,\nu_1, \mu_2 \,\nu_2 }^F(x_1p_1, x_2p_2; {\bf\Omega}; \as)
\nn \\
&\!\!\times&\vep_{(\lambda_1)}^{\,\mu_1}(x_1p_1) 
\;\left[ \vep_{(h_1)}^{\,\nu_1}(x_1p_1) \right]^*
\;\vep_{(\lambda_2)}^{\,\mu_2}(x_2p_2) 
\;\left[ \vep_{(h_2)}^{\,\nu_2}(x_2p_2)\right]^*
\;, 
\eeeq
\beq
\label{chel}
C_{g \,a}^{(\lambda_i, h_i)}(z_i;p_1,p_2,{\bf b};\as)
= \left[ \vep^{(\lambda_i)}_{\,\mu}(x_ip_i) \right]^*
\;C_{g \,a}^{\mu \,\nu}(z_i;p_1,p_2,{\bf b};\as) 
\;\vep^{(h_i)}_{\,\nu}(x_ip_i) \;, \;\; \quad i=1,2 \;\;,
\eeq
where $\vep^{(\lambda)}_{\,\mu}(p)$ denotes the polarization vector of a gluon
with on-shell momentum $p$
($p^2=0$) and helicity $\lambda$.

We recall that the gluon helicity vectors are not uniquely defined. Having chosen
the two helicity vectors $\vep^{(+)}_{\,\mu}(p)$ and $\vep^{(-)}_{\,\mu}(p)
=\left[ \vep^{(+)}_{\,\mu}(p) \right]^*$, there is still the freedom to change 
the helicity vector basis. For instance, the phase transformation
\beq
\label{poltran}
\vep^{(\lambda)}_{\,\mu}(p) \; \to \; 
{\widetilde \vep}^{\,\,(\lambda)}_{\,\mu}(p)
= e^{i \lambda \varphi_p} \;\vep^{(\lambda)}_{\,\mu}(p) \;\;,
\eeq
defines helicity states ${\widetilde \vep}^{\,\,(\lambda)}_{\,\mu}(p)$ that are
physically equivalent to $\vep^{\,\,(\lambda)}_{\,\mu}(p)$
(the phase $\varphi_p$ on the right-hand side of Eq.~(\ref{poltran}) can also
depend on the gluon momentum $p$). Note that Eq.~(\ref{whathel}) is invariant
under the transformation in Eq.~(\ref{poltran}), whereas the helicity tensors
$H_{(\lambda_1, h_1), (\lambda_2, h_2)}^F$ in Eq.~(\ref{hhel})
and $C_{g \,a}^{(\lambda, h)}$ in Eq.~(\ref{chel}) are not separately invariant. 
In any actual computation of $H_{(\lambda_1, h_1), (\lambda_2, h_2)}^F$ 
and $C_{g \,a}^{(\lambda, h)}$, the 
definition of the helicity basis has
to be clearly specified.  

Some general and important properties of the universal coefficient function 
$C_{g \,a}^{(\lambda, h)}$ do not depend on the specific definition of the
helicity basis. Owing to general relations between helicity vectors and using 
Eq.~(\ref{cggten}), we find 
that the helicity tensor
$C_{g \,a}^{(\lambda, h)}$ in Eq.~(\ref{chel}) has the following explicit
structure:
\beq
\label{chelCG}
C_{g \,a}^{(\lambda_i, h_i)}(z_i;p_1,p_2,{\bf b};\as)
= C_{g \,a}(z_i;\as) \;\delta^{\lambda_i, h_i} + G_{g \,a}(z_i;\as)
\;D^{(\lambda_i)}(p_i,{\bf b})
\;\delta^{\lambda_i, - h_i}
 \;\;, \;\; \quad i=1,2 \;\;,
\eeq
where the scalar coefficient functions $C_{g \,a}(z;\as)$ and $G_{g \,a}(z;\as)$
are those of Eq.~(\ref{cggten}), $\delta^{\lambda, h}$ is the customary Kronecker
symbol $(\delta^{+, +}=\delta^{-, -}=1 \;, \;\;\delta^{+, -}=\delta^{-, +}=0)$,
and
the helicity coefficients $D^{(\lambda)}(p_i,{\bf b})$,
\beq
\label{dlam}
D^{(\lambda)}(p_i,{\bf b}) = - \f{2}{{\bf b}^2} 
\;\left[ b \cdot \vep^{(- \lambda)}(x_ip_i)\right]^2 \;\;,
\;\; \quad i=1,2 \;\;,
\eeq
are pure phase factors that have the following explicit form:
\beeq
\label{dlam1}
D^{(\lambda)}(p_1,{\bf b}) &=& - \,e^{+ 2 i \lambda 
\;\left[\, \phi({\bf b}) - \varphi_1 \,\right] } \;\;, \\
\label{dlam2}
D^{(\lambda)}(p_2,{\bf b}) &=& - \,e^{- 2 i \lambda 
\;\left[\, \phi({\bf b}) + \varphi_2 \,\right] } \;\;.
\eeeq
Note that the helicity coefficients $D^{(\lambda_i)}(p_i,{\bf b})$ have a
distinctive dependence
on $\phi({\bf b})$ (which is the azimuthal angle of the impact parameter vector
$\bf b$), whereas the phases $\varphi_1$ and $\varphi_2$ simply depend on the
explicit
definition (to be specified) of the helicity vectors 
$\vep^{(\lambda)}_\mu(x_1p_1)$ and $\vep^{(\lambda)}_\mu(x_2p_2)$
(see Eq.~(\ref{poltran})). As already discussed, the dependence on 
$\varphi_1$ and $\varphi_2$ cancels in Eq.~(\ref{whathel}).

The helicity space representation (\ref{chelCG}) of the gluonic coefficient
tensors explicitly shows the presence of two components:
a helicity-conserving component and a helicity-flip component.
This two-component structure originates from the gluon collinear correlations
discussed in Sect.~\ref{sec:ggfus}. 
The helicity-conserving component leads to the \naive\
(in the case of gluon fusion processes) factorization formula in 
Eq.~(\ref{what}).
Inserting Eq.~(\ref{chelCG}) in Eq.~(\ref{whathel}), the helicity-flip component,
which is proportional to the coefficient function  $G_{g \,a}(z;\as)$
(see Eqs.~(\ref{gfexp})--(\ref{gq1})), obviously produces non-trivial helicity
(spin) correlations in the process-dependent factor $H^F$.

The helicity-flip coefficients in Eqs.~(\ref{dlam1}) and (\ref{dlam2}) can be
rewritten as follows
\beeq
\label{dlam1b}
D^{(\lambda)}(p_1,{\bf b}) &=& - \,e^{+ 2 i \lambda 
\;\left[\, \phi({\bf b}\cdot \bqt) - \,\varphi_1 \,\right] } 
\;e^{+ 2 i \lambda \phi({\bqt}) } = \,e^{+ 2 i \lambda \phi({\bf b}\cdot{\bqt}) }
\;D^{(\lambda)}(p_1,{\bqt})
\;\;, \\
\label{dlam2b}
D^{(\lambda)}(p_2,{\bf b}) &=& - \,e^{- 2 i \lambda 
\;\left[\, \phi({\bf b}\cdot \bqt) + \varphi_2 \,\right] } 
\;e^{- 2 i \lambda \phi({\bqt}) } = \,e^{- 2 i \lambda \phi({\bf b}\cdot{\bqt}) }
\;D^{(\lambda)}(p_2,{\bqt})
\;\;,
\eeeq
where $\phi({\bqt})$ is the azimuthal angle of ${\bqt}$ (in the centre--of--mass
frame of the collision), and $\phi({\bf b}\cdot{\bqt}) \equiv 
\phi({\bf b}) - \phi({\bqt})$ is the relative angle between 
${\bqt}$ and the impact parameter
(i.e. ${\bf b}\cdot{\bqt}= b \,\qt \,\cos \phi({\bf b}\cdot{\bqt})\,$).
Using Eqs.~(\ref{whathel}) and (\ref{chelCG}), the helicity-flip phase factors
in Eqs.~(\ref{dlam1b}) and (\ref{dlam2b}) eventually enter the cross section
formula in Eq.~(\ref{qtycrossgg}). Since $\phi({\bf b}\cdot{\bqt})$ is simply the
angular integration variable of the Fourier transformation in 
Eq.~(\ref{qtycrossgg}), the dependence on $\phi({\bqt})$ of the $\bqt$ cross
section $\left[ d\sigma_F \right]$ is directly determined by the phase factors
$\exp\left(\pm 2i \lambda \,\phi({\bqt}) \right)$ in 
Eqs.~(\ref{dlam1b}) and (\ref{dlam2b}). Therefore, the helicity-flip component
of the gluonic helicity tensor in Eq.~(\ref{chelCG}) controls both the spin
correlations and the azimuthal correlations of the ${\bqt}$ cross section
in the small-$\qt$ region.

In Sect.~\ref{sec:ggfus} we have discussed the case of SM Higgs 
boson production by gluon
fusion.
The corresponding helicity tensor 
$H_{(\lambda_1, h_1), (\lambda_2, h_2)}^{F=H}$ is obtained by using 
Eqs.~(\ref{hh}) and (\ref{hhel}). We have:
\beq
\label{hhhel}
H_{(\lambda_1, h_1), (\lambda_2, h_2)}^{F=H}(x_1p_1,x_2p_2;\as(M^2))
= \f{1}{2} \;H_{g}^{F=H}(\as(M^2)) 
\;\delta^{\lambda_1,\lambda_2} \;\delta^{h_1,h_2}  \;
e^{i (\lambda_1 - h_1) (\varphi_1 + \varphi_2)}
\;\;,
\eeq
where we have used the relation
\beq
\vep^{(\lambda_1)}_{\,\mu_1}(x_1p_1)
\;\left( g^{\mu_1 \mu_2} - \f{p_{2}^{\mu_1} p_{1}^{\mu_2}}{p_1 \cdot p_2}\right)
\;\vep^{(\lambda_2)}_{\,\mu_2}(x_2p_2) = 
- \,e^{i \lambda_1 (\varphi_1 + \varphi_2)} \;\delta^{\lambda_1,\lambda_2}
\;\;.
\eeq
In Eq.~(\ref{hhhel}), the constraint $\lambda_1=\lambda_2$ (and $h_1=h_2$)
obviously originates from helicity conservation for the production of a boson with
spin $0$.
Inserting Eqs.~(\ref{chelCG}),
(\ref{dlam1}), (\ref{dlam2}) and (\ref{hhhel}) in  Eq.~(\ref{whathel}), we
reobtain the
result in Eq.~(\ref{whath}). Terms proportional to $C(z_1) G(z_2)$ are absent
from Eq.~(\ref{whath}), since helicity conservation in the hard-process factor
of Eq.~(\ref{hhhel}) forbids contributions 
that are produced by a single helicity flip.

We comment on some of the results of 
Refs.~[\ref{Nadolsky:2007ba}, \ref{Balazs:2007hr}]. 
Studying diphoton production, the authors of Ref.~[\ref{Nadolsky:2007ba}]
made the following important observation: the description of the 
small-$\qt$ behaviour of the diphoton transverse-momentum cross section
requires the introduction of new logarithmically-enhanced 
spin-flip contributions (which affect the azimuthal angle dependence of the
produced diphoton system), through a mechanism that is unique to
gluon scattering. This general observation was based [\ref{Nadolsky:2007ba}]
on the computation of the first-order QCD radiative corrections to the
lowest-order gluon fusion process $gg \to \gamma \gamma$. Expanding our resummed
formulae (see, e.g., Eqs.~(\ref{qtycrossgg}), (\ref{whathel}) and 
(\ref{chelCG})) to the first order in $\as$, we find agreement with the structure
of the diphoton first-order results of Ref.~[\ref{Nadolsky:2007ba}]. To be
precise, we `almost' agree with those first-order results: our expression
contains a term proportional to $\sum_{a=g,q,{\bar q}} \,G^{(1)}_{g \,a}(z)
\,f_{a/h}(x/z)$, whereas the expression in Ref.~[\ref{Nadolsky:2007ba}] includes
only the contribution $G^{(1)}_{g \,g}(z) \,f_{g/h}(x/z)$
(the function $P^{\,\prime}_{g/g}(z)$ of Ref.~[\ref{Nadolsky:2007ba}]
is related to $G^{(1)}_{g \,g}(z)$ in Eq.~(\ref{gg1}); namely, 
$P^{\,\prime}_{g/g}(z)= 2 \,G^{(1)}_{g \,g}(z)\,$). The authors of 
Ref.~[\ref{Nadolsky:2007ba}] also proposed an all-order generalization of their
first-order results. The generalization (which is expressed in the
Collins--Soper frame [\ref{Collins:1977iv}] of the diphoton pair)
is achieved by introducing a new helicity-flip unintegrated parton density
of the colliding hadrons. This unintegrated (transverse-momentum-dependent)
parton density, or, more precisely, its Fourier transform to $\bf b$ space
is denoted by $\delta^{\lambda, -h} \;{\cal P}^{\,\prime}_{g/h}(x,{\bf b})$,
where  ${\cal P}^{\,\prime}_{g/h}(x,{\bf b})$ does not depend on the gluon
helicity state $\lambda = -h$. The general dependence of 
${\cal P}^{\,\prime}_{g/h}(x,{\bf b})$ on $\bf b$ and the possible relation
between ${\cal P}^{\,\prime}_{g/h}(x,{\bf b})$ and the customary parton
densities are not specified in Ref.~[\ref{Nadolsky:2007ba}]
(apart from the relation with $f_{g/h}$ that can be inferred from the
first-order calculation therein). Our resummation formulae also contain one
helicity-flip component for each of the colliding hadrons. This component is
proportional to the following expression:
\beq
\label{helflip}
\delta^{\lambda, -h} \;\int_x^1 \f{dz}{z} \;\sum_{a=g,q,{\bar q}} 
\,G_{g \,a}(z; \as(b_0^2/b^2)) \;f_{a/h_i}(x/z, b_0^2/b^2 )
\;D^{(\lambda)}(p_i,{\bf b}) \;\;, \;\;\quad i=1,2 \;\;,
\eeq
which is directly related to the parton densities $f_{a/h}$ 
through the longitudinal-momentum convolution with the 
perturbatively-calculable gluonic coefficient functions 
$G_{g \,a}(z; \as)$ $(a=g,q,{\bar q})$. Note also that the expression in
Eq.~(\ref{helflip}) explicitly depends on the gluon helicity $\lambda$ through
the helicity coefficients $D^{(\lambda)}(p_i,{\bf b})$ $(i=1,2)$. The 
helicity coefficients introduce a definite and non-trivial (though, simple)
functional dependence on the azimuthal angle $\phi({\bf b})$ 
(see Eqs.~(\ref{dlam1}) and (\ref{dlam2})) or, equivalently,
on the relative azimuthal angle $\phi({\bf b}\cdot \bqt)$
(see Eqs.~(\ref{dlam1b}) and (\ref{dlam2b})).
This functional dependence eventually produces definite {\em coherent}
correlations (see, e.g., Eq.~(\ref{whath}) and, more generally, 
Eqs.~(\ref{hGhel}), (\ref{fbGGhel}) and (\ref{qtycrossphi}) in 
Sects.~\ref{sec:azav} and \ref{sec:FB}) between the
helicities, $\lambda_1$ and $\lambda_2$, of the two colliding gluons.
Owing to this helicity dependence of the helicity-flip components in
Eq.~(\ref{helflip}), by using our gluon fusion resummed formulae we are not able
to reproduce the diphoton results of Ref.~[\ref{Nadolsky:2007ba}]
(e.g., the structure of
Eqs.~(27), (30) and (38) in Ref.~[\ref{Nadolsky:2007ba}]): the
differences show up starting from contributions of relative order $\as^2$
with respect to the lowest-order process $gg \to \gamma \gamma$.
Few additional comments about this point are included in the final part of
Sect.~\ref{sec:FB}.

\section{Azimuthally-averaged cross sections in gluon fusion\\
 processes}
\label{sec:azav}

Starting from the general $\bqt$ cross section in Eq.~(\ref{diffxs}),
we can consider its average over $\phi(\bqt)$, at fixed values of the 
additional
kinematical variables $\bf \Omega$ of the final-state system $F$.
We thus define the following azimuthally-averaged cross section:
\beeq
\label{avdiffxs}
{\big \langle} \;\f{d\sigma_F}{d^2{\bqt} \;dM^2 \;dy \;d{\bf\Omega}}  
\;{\big \rangle}_{\phi} &\equiv&
\int_0^{2\pi} \f{d\phi(\bqt)}{2\pi} 
\;\f{d\sigma_F}{d^2{\bqt} \;dM^2 \;dy \;d{\bf\Omega}} \nn \\
&=&
\f{1}{\pi} \;\f{d\sigma_F}{dq_T^2 \;dM^2 \;dy \;d{\bf\Omega}} 
\,(p_1, p_2;\qt^2,M,y,
{\bf\Omega} )
\;\;.
\eeeq
Following the shorthand notation of Eq.~(\ref{singsig}),
the singular component (in the small-$\qt$ region) of the cross section 
(\ref{avdiffxs}) is denoted as $\left[ d\sigma_F \,\right]_{\phi}$.
Then we consider the decomposition in Eq.~(\ref{qgdec}). As recalled at the end
of Sect.~\ref{sec:revqqbar}, in the case of $q{\bar q}$ 
annihilation subprocesses,
small-$\qt$ resummation does not produce any azimuthal correlations and,
therefore,
the corresponding $\left[ d\sigma_F \,\right]^{(q{\bar q}-{\rm ann.})}$
and $\left[ d\sigma_F \,\right]_{\phi}^{(q{\bar q}-{\rm ann.})}$ 
are obtained by the same resummation
formula (see Eqs.~(\ref{qtycross}) and (\ref{what}) with $c=q, {\bar q}$).
In the case of gluon fusion processes, 
the resummation formula for the azimuthally-averaged cross
section is
obtained from Eq.~(\ref{qtycrossgg}) by replacing 
\beq
\left[ d\sigma_F \right]^{(g-{\rm fus.})} \; \to \;
\left[ d\sigma_F \right]_{\phi}^{(g-{\rm fus.})} \;\;,
\eeq
and by performing the following replacement
\beq
\left[ H^F C_1 C_2 \right]_{gg;a_1a_2} \; \to \;
\left[ H^F C_1 C_2 \right]^{\phi}_{gg;a_1a_2}
\eeq
in the integrand on the right-hand side;
the two-dimensional Fourier transformation con be replaced by the
Bessel transformation as in Eq.~(\ref{fourbes}).
Therefore, the resummation formula for the azimuthally-averaged cross
section is:
\beeq
\label{qtycrossggphi}
&&\left[ d\sigma_F \right]_{\phi}^{(g-{\rm fus.})}
 =\f{M^2}{s} \;
 \left[ d\sigma_{gg, \,F}^{(0)} \right]
\int_0^{+\infty} \f{db}{2\pi} \;b \;\,J_0(b\qt) \;\,S_g(M,b)\nn \\
&& \;\;\;\; \times \;
\sum_{a_1,a_2} \,
\int_{x_1}^1 \f{dz_1}{z_1} \,\int_{x_2}^1 \f{dz_2}{z_2} 
\; \left[ H^F C_1 C_2 \right]_{gg;a_1a_2}^{\phi}
\;f_{a_1/h_1}(x_1/z_1,b_0^2/b^2)
\;f_{a_2/h_2}(x_2/z_2,b_0^2/b^2) \;
\;, 
\eeeq
where
the integrand factor $\left[ H^F C_1 C_2 \right]^{\phi}_{gg;a_1a_2}$
has the following explicit expression: 
\beeq
\label{whatggphi}
\!\!\!\!\!\! \left[ H^F C_1 C_2 \right]^{\phi}_{gg;a_1a_2}
&=& H_{g}^F(x_1p_1, x_2p_2; {\bf\Omega}; \as(M^2)) 
\; C_{g \,a_1}(z_1;\as(b_0^2/b^2)) 
\;\, C_{g \,a_2}(z_2;\as(b_0^2/b^2)) 
\nn \\
&+& H_{G}^F(x_1p_1, x_2p_2; {\bf\Omega}; \as(M^2))
\; G_{g \,a_1}(z_1;\as(b_0^2/b^2)) 
\;\, G_{g \,a_2}(z_2;\as(b_0^2/b^2)) 
\;\;. 
\eeeq
This result can easily be obtained by using Eqs.~(\ref{whathel}), 
(\ref{chelCG}), (\ref{dlam1b}) and (\ref{dlam2b})
and the following
elementary integral:
\beq
\int_0^{2\pi} \f{d\phi(\bqt)}{2\pi} \;e^{\pm \,i \,n \,\phi(\bqt)} = \delta^{\,n, 0} 
\;\;,
 \;\; \quad n=0,1,2,\dots. \;\;\;.
\eeq

The process-dependent factor $H_{g}^F$ on the right-hand side of 
Eq.~(\ref{whatggphi}) is defined in Eq.~(\ref{hgg}). Its equivalent
representation in helicity space is
\beq
\label{hgghel}
H_{g}^F(x_1p_1, x_2p_2; {\bf\Omega}; \as) = \sum_{\lambda_1, \lambda_2}
\; H_{(\lambda_1, \lambda_1), (\lambda_2,\lambda_2)}^F(x_1p_1, x_2p_2; 
{\bf\Omega}; \as) \;\;.
\eeq
The spin-correlated hard-scattering factor $H_{G}^F$ has the following
expression:
\beeq
\label{hGhel}
H_{G}^F(x_1p_1, x_2p_2; {\bf\Omega}; \as) \!\!\!&=&\!\!\sum_{\lambda}
\; H_{(\lambda, - \lambda), (\lambda, - \lambda)}^F(x_1p_1, x_2p_2; 
{\bf\Omega}; \as) \;\,e^{- 2 i \lambda (\varphi_1+ \varphi_2)} \\
\label{hG}
\!\!\!&=&\!\! \;
H_{\mu_1 \,\nu_1, \mu_2 \,\nu_2 }^F(x_1p_1, x_2p_2; {\bf\Omega}; \as) 
\;\,d_{(4)}^{\mu_1 \,\nu_1, \mu_2 \,\nu_2}(p_1,p_2) \;\;,
\eeeq
where the 4th-rank tensor $d_{(4)}^{\mu_1 \,\nu_1, \mu_2 \,\nu_2}$ is
\beq
\label{d4ten}
d_{(4)}^{\,\mu_1 \,\nu_1, \mu_2 \,\nu_2}(p_1,p_2) = \f{1}{2} \;
\Bigl[ 
\;d^{\,\mu_1 \mu_2} \,d^{\,\nu_1 \nu_2} 
+ d^{\,\mu_1 \nu_2} \,d^{\,\mu_2 \nu_1}
- d^{\,\mu_1 \nu_1} \,d^{\,\mu_2 \nu_2} \,
\Bigr] \;,
\eeq
with $d^{\mu \nu}=d^{\mu \nu}(p_1,p_2)$ (see Eq.~(\ref{dten})).
From the helicity space representation (\ref{hGhel}) of the
process-dependent factor $H_{G}^F$, we see that both gluons (with momenta 
$x_1p_1$ and  $x_2p_2$) undergo a helicity flip (i.e., $\lambda_1= - h_1$
and $\lambda_2= - h_2$); moreover, we note that the two helicity flips are
{\em correlated}: they occur {\em coherently},
with the constraint $\lambda_1=\lambda_2$.

As already anticipated in Sect.~\ref{sec:ggfus}, in gluon fusion processes
the \naive\ resummation factor of 
Eq.~(\ref{what}) is not correct even in the case of azimuthally-averaged
cross sections. The correct resummation factor in given in 
Eq.~(\ref{whatggphi}). Comparing Eqs.~(\ref{what}) and 
(\ref{whatggphi}), we note the presence of the additional gluonic coefficient 
function $G_{g \,a}(z;\as)$ and of the corresponding hard-scattering factor
$H_{G}^F$ (which, in general\footnote{In the specific case of SM Higgs boson
production, we have $H_{G}^{F=H}=H_{g}^{F=H}$ 
(see Eq.~(\ref{whath})).}, differs from the spin-uncorrelated factor 
$H_{g}^F$).

\section{Fourier vs. Bessel transformations and azimuthal 
correlations of the $\bqt$ cross section}
\label{sec:FB}

Unlike the case of $q{\bar q}$ annihilation,
the gluon fusion resummation factor 
$\left[ H^F C_1 C_2 \right]$ in Eqs.~(\ref{whatgg}) or 
(\ref{whathel}) depends on the azimuthal angle $\phi({\bf b})$ of the impact
parameter vector ${\bf b}$. Therefore, in the resummation formula
(\ref{qtycrossgg}) we cannot simply use the relation (\ref{fourbes})
to perform the azimuthal integration involved in the Fourier transformation from
${\bf b}$ space to $\bqt$ space.
Nonetheless, the dependence of $\left[ H^F C_1 C_2 \right]_{gg;a_1a_2}$
on $\phi({\bf b})$ is fully specified at arbitrary perturbative orders. 
Moreover, this dependence is sufficiently simple to be handled in explicit form.
As shown below, the azimuthal integration involved in the Fourier 
transformation can explicitly be performed. The two-dimensional Fourier 
transformation is thus replaced by one-dimensional Bessel transformations. The
weight functions
of these one-dimensional transformations are the $0$th-order Bessel function
$J_0(b\qt)$ and higher-order Bessel functions, such as, the $2$nd-order and
$4$th-order functions $J_2(b\qt)$ and $J_4(b\qt)$.

To present the relation between Fourier and Bessel transformations, we first
consider the gluon fusion resummation formula in helicity space. In this
formulation,
the $\phi({\bf b})$ dependence of the integrand factor 
$\left[ H^F C_1 C_2 \right]$ (see Eqs.~(\ref{whathel}) and (\ref{chelCG}))
is given by the helicity coefficients
$D^{(\lambda)}(p_i,{\bf b})$ in Eqs.~(\ref{dlam1})--(\ref{dlam2b}). We have to
perform the Fourier transformation of contributions that are linear and quadratic
with respect to these helicity coefficients. The explicit results of the
corresponding integration over $\phi({\bf b})$ are the following:
\beeq
\label{fbGhel}
\int \f{d^2{\bf b}}{2\pi} \;\, e^{i {\bf b}\cdot \bqt} 
\;D^{(\lambda)}(p_i,{\bf b})\;\;F({\bf b}^2)
&=& D^{(\lambda)}(p_i,{\bqt}) \;\int \f{d^2{\bf b}}{2\pi} 
\;\, e^{i {\bf b}\cdot \bqt} \;\, e^{\pm 2 i \lambda \,\phi({\bf b}\cdot \bqt)}
\;\;F({\bf b}^2) \nn \\
&=& - \;D^{(\lambda)}(p_i,{\bqt})
\;\int_0^{+\infty} \,db \;b  \;J_2(b q_T) \;\;F(b^2) \;\;,
\eeeq
\beeq
\label{fbGGhel}
&&\int \f{d^2{\bf b}}{2\pi} \;\, e^{i {\bf b}\cdot \bqt} 
\;D^{(\lambda_1)}(p_1,{\bf b}) \;D^{(\lambda_2)}(p_2,{\bf b})\;\;F({\bf b}^2) 
\nn \\
&& = D^{(\lambda_1)}(p_1,{\bqt}) \;D^{(\lambda_2)}(p_2,{\bqt})
\;\int \f{d^2{\bf b}}{2\pi} 
\;\, e^{i {\bf b}\cdot \bqt} 
\;\, e^{+ 2 i (\lambda_1 - \lambda_2) \,\phi({\bf b}\cdot \bqt)}
\;\;F({\bf b}^2) \nn \\
&&= D^{(\lambda_1)}(p_1,{\bqt}) \;D^{(\lambda_2)}(p_2,{\bqt})
\;\int_0^{+\infty} \,db \;b 
\left[ \,\delta^{\lambda_1, \,\lambda_2} \;J_0(b q_T)
+ \delta^{\lambda_1, \,-\lambda_2} \;J_4(b q_T) \,
\right] \;F(b^2)
\nn \\
&&= \;\delta^{\lambda_1, \,\lambda_2} 
\;\, e^{- 2 i \lambda_1 \,(\varphi_1 + \varphi_2)}
\;\int_0^{+\infty} \,db \;b  \;J_0(b q_T) \;F(b^2) \nn \\
&&+ \;\;\delta^{\lambda_1, \,-\lambda_2} 
\,\;D^{(\lambda_1)}(p_1,{\bqt}) \;D^{(\lambda_2)}(p_2,{\bqt})
\;\int_0^{+\infty} \,db \;b  \;J_4(b q_T) \;F(b^2)\;\;,
\eeeq
where $F({\bf b}^2)$ denotes a generic function of ${\bf b}^2$.
The results in Eqs.~(\ref{fbGhel}) and (\ref{fbGGhel}) are straightforwardly
obtained by simply using the following integral representation of the Bessel
function~$J_{2n}$:
\beq
J_{2n}(x)= (-1)^n \;
\int_0^{2\pi} \f{d\phi}{2\pi} \;e^{+ i x \cos\phi} \;e^{\pm i 2 n \phi} 
\;\;,
 \;\; \quad n=0,1,2,\dots \;\;\;,
\eeq
where the variable $x$ is a real number.

Considering the formulation in terms of Lorentz tensors
(see Eqs.~(\ref{whatgg}) and (\ref{cggten})), the dependence on $\phi({\bf b})$ 
is produced by the tensor $D^{\,\mu \nu}(p_1,p_2;{\bf b})$
in Eq.~(\ref{dbten}). The results analogous to those in 
Eqs.~(\ref{fbGhel}) and (\ref{fbGGhel}) are:
\beq
\label{fbG}
\int \f{d^2{\bf b}}{2\pi} \;\, e^{i {\bf b}\cdot \bqt} 
\;D^{\,\mu \nu}(p_1,p_2;{\bf b})\;\;F({\bf b}^2)
= - \;D^{\,\mu \nu}(p_1,p_2;{\bqt})
\;\int_0^{+\infty} \,db \;b  \;J_2(b q_T) \;\;F(b^2) \;\;,
\eeq
\beeq
\label{fbGG}
&&\int \f{d^2{\bf b}}{2\pi} \;\, e^{i {\bf b}\cdot \bqt} 
\;D^{\,\mu_1 \nu_1}(p_1,p_2;{\bf b}) \;D^{\,\mu_2 \nu_2}(p_1,p_2;{\bf b})
\;\;F({\bf b}^2) \nn \\
&&= \;\;d_{(4)}^{\,\mu_1 \,\nu_1,\, \mu_2 \,\nu_2}(p_1,p_2) 
\;\int_0^{+\infty} \,db \;b  \;J_0(b q_T) \;\;F(b^2) \nn \\
&&+ \;\;\;D_{(4)}^{\,\mu_1 \,\nu_1,\, \mu_2 \,\nu_2}(p_1,p_2;\bqt)
\;\int_0^{+\infty} \,db \;b  \;J_4(b q_T) \;\;F(b^2)
\;\;,
\eeeq
where $d_{(4)}^{\,\mu_1 \,\nu_1,\, \mu_2 \,\nu_2}$ is given in Eq.~(\ref{d4ten}),
and the $4$th-rank tensor $D_{(4)}^{\,\mu_1 \,\nu_1,\, \mu_2 \,\nu_2}$ is
\beq
\label{d44ten}
D_{(4)}^{\,\mu_1 \,\nu_1,\, \mu_2 \,\nu_2}(p_1,p_2;\bqt) = 
D^{\,\mu_1 \nu_1}(p_1,p_2;{\bqt}) \;D^{\,\mu_2 \nu_2}(p_1,p_2;{\bqt})
- d_{(4)}^{\,\mu_1 \,\nu_1, \mu_2 \,\nu_2}(p_1,p_2)  \;\;.
\eeq
The derivation of Eqs.~(\ref{fbG}) and (\ref{fbGG}) is similar to the derivation
of Eqs.~(\ref{fbGhel}) and (\ref{fbGGhel}), apart from few additional algebraic
manipulations.

We note that the three Bessel functions $J_0, J_2$ and $J_4$ are not independent.
Owing to general recursion relations between Bessel functions, we have:
\beq
J_4(x) =\f{4 \,(6-x^2)}{x^2} \,J_2(x) - 3 \,J_0(x) \;\;.
\eeq
This relation can be used to express the results in Eqs.~(\ref{fbGhel}),
(\ref{fbGGhel}), (\ref{fbG}) and (\ref{fbGG}) in terms of two (rather than three)
Bessel functions.

The representation in terms of Bessel transformations offers a technical 
simplification of the resummation formula, 
since the two-dimensional Fourier transformation is replaced by 
one-dimen\-sional transformations. Moreover, in Eqs.~(\ref{fbGhel}) and
(\ref{fbGGhel}) (or, equivalently, in Eqs.~(\ref{fbG}) and (\ref{fbGG}))
the dependence on the azimuthal angle $\phi(\bqt)$ is fully and explicitly
factorized with respect to the dependence on the magnitude, $\qt$, of the
transverse momentum (the dependence on $\qt$ is produced by the integration over
$b$). Therefore, we can express the gluon fusion resummation formula in a form
that manifestly exhibits the functional dependence on  
$\phi(\bqt)$ of the $\bqt$ cross section, at small values of $\qt$.

To be explicit, we insert Eqs.~(\ref{fbGhel}) and
(\ref{fbGGhel}) (or, Eqs.~(\ref{fbG}) and (\ref{fbGG})) 
in the resummation formula
(\ref{qtycrossgg}), and we obtain:
\beeq
\left[ d\sigma_F \right]^{(g-{\rm fus.})} &=&
\left[ d\sigma_F \right]_{\phi}^{(g-{\rm fus.})} \nn \\
&+&
\left[ d\sigma_F \right]_{C_1G_2} 
\;\left[ H^F(\phi(\bqt)) \right]_{C_1G_2}+
\;\left[ d\sigma_F \right]_{G_1C_2}
\;\left[ H^F(\phi(\bqt)) \right]_{G_1C_2} \nn \\
&+& 
\label{qtycrossphi}
\left[ d\sigma_F \right]_{GG} 
\;\left[ H^F(\phi(\bqt)) \right]_{GG}
\;\;.
\eeeq
The notation $\left[ H^F(\phi(\bqt))\right]_{I}$ 
(rather than simply
$\left[ H^F \,\right]_{I}\,$) remarks that each of these factors 
(the subscript $I$ is $I=C_1G_2, G_1C_2$ or $GG$) 
depends on $\bqt$ 
only through
its azimuthal angle $\phi(\bqt)$
(i.e., these factors do not depend on the magnitude of 
$\bqt$). All the other terms on the right-hand side of Eq.~(\ref{qtycrossphi}) 
depend on $\qt$, but they are independent of $\phi(\bqt)$.

In Eq.~(\ref{qtycrossphi}), the gluon fusion cross section 
$\left[ d\sigma_F \right]^{(g-{\rm fus.})}$
is partitioned into several contributions. The first contribution
is equal to $\left[ d\sigma_F \right]_{\phi}^{(g-{\rm fus.})}$,
the azimuthally-averaged cross section. Since the cross section
$\left[ d\sigma_F \right]^{(g-{\rm fus.})}$ is evaluated at fixed values of 
$\phi(\bqt)$, our notation is imprecise. The notation actually means that the
first contribution on the right-hand side of Eq.~(\ref{qtycrossphi})
is explicitly given by the expression presented on the right-hand side 
of Eq.~(\ref{qtycrossggphi}); this expression coincides with the expression of
the azimuthal average over $\phi(\bqt)$ of 
$\left[ d\sigma_F \right]^{(g-{\rm fus.})}$.
The other cross section contributions, $\left[ d\sigma_F \right]_I$,
in Eq.~(\ref{qtycrossphi}) have the following form:
\beeq
\label{qtycrossI}
&&\left[ d\sigma_F \right]_{I}
 =\f{M^2}{s} \;
 \left[ d\sigma_{gg, \,F}^{(0)} \right]
\int_0^{+\infty} \f{db}{2\pi} \;b \;\,S_g(M,b)\nn \\
&& \;\;\;\; \times \;
\sum_{a_1,a_2} \,
\int_{x_1}^1 \f{dz_1}{z_1} \,\int_{x_2}^1 \f{dz_2}{z_2} 
\; \left[ J C_1 C_2 \right]_{gg;a_1a_2}^I
\;f_{a_1/h_1}(x_1/z_1,b_0^2/b^2)
\;f_{a_2/h_2}(x_2/z_2,b_0^2/b^2) \;
\;, 
\eeeq
where the integrand factors denoted by
$\left[ J C_1 C_2 \right]^I$ are given by the following explicit expressions:
\beq
\label{jcg}
\left[ J C_1 C_2 \right]_{gg;a_1a_2}^{C_1G_2} = J_2(b\qt)
\;\,C_{g \,a_1}(z_1;\as(b_0^2/b^2)) 
\;\, G_{g \,a_2}(z_2;\as(b_0^2/b^2)) \;\;,
\eeq
\beq
\label{jgc}
\left[ J C_1 C_2 \right]_{gg;a_1a_2}^{G_1C_2} = J_2(b\qt)
\;\,G_{g \,a_1}(z_1;\as(b_0^2/b^2)) 
\;\, C_{g \,a_2}(z_2;\as(b_0^2/b^2)) \;\;,
\eeq
\beq
\label{jgg}
\left[ J C_1 C_2 \right]_{gg;a_1a_2}^{GG} = J_4(b\qt)
\;\,G_{g \,a_1}(z_1;\as(b_0^2/b^2)) 
\;\, G_{g \,a_2}(z_2;\as(b_0^2/b^2)) \;\;.
\eeq
The hard-scattering factors $\left[ H^F(\phi(\bqt)) \right]_{I}$
are process dependent; they are
\beeq
\label{hcgl}
\left[ H^F(\phi(\bqt)) \right]_{C_1G_2} &=& - \,
H_{\mu_1 \,\nu_1, \mu_2 \,\nu_2 }^F(x_1p_1, x_2p_2; {\bf\Omega}; \as(M^2))
\;d^{\,\mu_1 \,\nu_1}(p_1,p_2)
\;D^{\,\mu_2 \,\nu_2}(p_1,p_2;\bqt) \\
&=& \sum_{\lambda_1, \,\lambda_2}
\; H_{(\lambda_1, \lambda_1), (\lambda_2, - \lambda_2)}^F(x_1p_1, x_2p_2; 
{\bf\Omega}; \as(M^2)) \;\,e^{- 2 i \lambda_2 \,(\varphi_2+ \phi(\bqt))}
\nn \\
\label{hcg}
&=& \cos\bigl(2\phi(\bqt)\bigr) \;
\sum_{\lambda_1, \,\lambda_2}
\; H_{(\lambda_1, \lambda_1), (\lambda_2, - \lambda_2)}^F(x_1p_1, x_2p_2; 
{\bf\Omega}; \as(M^2)) \;\,e^{- 2 i \lambda_2 \,\varphi_2} 
\\
&-& \sin\bigl(2\phi(\bqt)\bigr) \;
\sum_{\lambda_1, \,\lambda_2} \;i \,\lambda_2
\; H_{(\lambda_1, \lambda_1), (\lambda_2, - \lambda_2)}^F(x_1p_1, x_2p_2; 
{\bf\Omega}; \as(M^2)) \;\,e^{- 2 i \lambda_2 \,\varphi_2}
\;, \nn
\eeeq
\beeq
\left[ H^F(\phi(\bqt)) \right]_{G_1C_2} &=& - \,
H_{\mu_1 \,\nu_1, \mu_2 \,\nu_2 }^F(x_1p_1, x_2p_2; {\bf\Omega}; \as(M^2))
\;D^{\,\mu_1 \,\nu_1}(p_1,p_2;\bqt)
\;d^{\,\mu_2 \,\nu_2}(p_1,p_2) \\
&=& \sum_{\lambda_1, \,\lambda_2}
\; H_{(\lambda_1, - \lambda_1), (\lambda_2, \lambda_2)}^F(x_1p_1, x_2p_2; 
{\bf\Omega}; \as(M^2)) \;\,e^{- 2 i \lambda_1 \,(\varphi_1- \phi(\bqt))}
\nn \\
\label{hgc}
&=& \cos\bigl(2\phi(\bqt)\bigr) \;
\sum_{\lambda_1, \,\lambda_2}
\; H_{(\lambda_1, - \lambda_1), (\lambda_2, \lambda_2)}^F(x_1p_1, x_2p_2; 
{\bf\Omega}; \as(M^2)) \;\,e^{- 2 i \lambda_1 \,\varphi_1}
\\
&+& \sin\bigl(2\phi(\bqt)\bigr) \;
\sum_{\lambda_1, \,\lambda_2} \;i \,\lambda_1
\; H_{(\lambda_1, - \lambda_1), (\lambda_2, \lambda_2)}^F(x_1p_1, x_2p_2; 
{\bf\Omega}; \as(M^2)) \;\,e^{- 2 i \lambda_1 \,\varphi_1}
\;, \nn
\eeeq
\beeq
\left[ H^F(\phi(\bqt)) \right]_{GG} &=&
H_{\mu_1 \,\nu_1, \mu_2 \,\nu_2 }^F(x_1p_1, x_2p_2; {\bf\Omega}; \as(M^2))
\;\,D_{(4)}^{\,\mu_1 \,\nu_1,\, \mu_2 \,\nu_2}(p_1,p_2;\bqt) \\
&=& \sum_{\lambda}
\; H_{(\lambda, \,- \lambda), (-\lambda, \,\lambda)}^F(x_1p_1, x_2p_2; 
{\bf\Omega}; \as(M^2)) \;\,e^{- 2 i \lambda \,(\varphi_1- \varphi_2 
- 2 \phi(\bqt))}
\nn \\
\label{hGG}
&=& \cos\bigl(4\phi(\bqt)\bigr) \;
\sum_{\lambda}
\; H_{(\lambda, \,- \lambda), (-\lambda, \,\lambda)}^F(x_1p_1, x_2p_2; 
{\bf\Omega}; \as(M^2)) \;\,e^{- 2 i \lambda \,(\varphi_1- \varphi_2 )}
\\
&+& \sin\bigl(4\phi(\bqt)\bigr) \;
\sum_{\lambda} \;i \,\lambda
\; H_{(\lambda, \,- \lambda), (-\lambda, \,\lambda)}^F(x_1p_1, x_2p_2; 
{\bf\Omega}; \as(M^2)) \;\,e^{- 2 i \lambda \,(\varphi_1- \varphi_2 )}
\;. \nn
\eeeq
The factors $\left[ H^F(\phi(\bqt)) \right]_{C_1G_2}$ and
$\left[ H^F(\phi(\bqt)) \right]_{G_1C_2}$ involve a single helicity flip
(see Eqs.~(\ref{hcg}) and (\ref{hgc})). From the helicity space representation 
(\ref{hGG}) of the factor $\left[ H^F(\phi(\bqt)) \right]_{GG}$, 
we see that both gluons
(with momenta $x_1 p_1$ and $x_2 p_2$) undergo a helicity flip, and the two 
helicity flips are {\em correlated} by the constraint $\lambda_1= - \lambda_2$.
We recall that the double helicity flip with $\lambda_1= \lambda_2$ leads to the
factor $H_{G}^F$ (see Eq.~(\ref{hGhel})) that enters the cross section
contribution  
$\left[ d\sigma_F \right]_{\phi}^{(g-{\rm fus.})}$
(see Eq.~(\ref{whatggphi})).

By direct inspection of Eqs.~(\ref{qtycrossphi}), (\ref{hcg}), 
(\ref{hgc}) and (\ref{hGG}), we note that the $\phi(\bqt)$ azimuthal dependence
of the logarithmically-enhanced terms at small $\qt$ is fully determined by the
general structure of the transverse-momentum resummation formula. The $\bqt$
cross section 
$\left[ d\sigma_F \right]^{(g-{\rm fus.})}$ contains a contribution that is
independent of $\phi(\bqt)$ (namely, the term
$\left[ d\sigma_F \right]_{\phi}^{(g-{\rm fus.})}$) plus a
linear combination of
the four trigonometric functions
$\cos\bigl(2\phi(\bqt)\bigr), \, \sin\bigl(2\phi(\bqt)\bigr), \,
\cos\bigl(4\phi(\bqt)\bigr)$ and $\sin\bigl(4\phi(\bqt)\bigr)$.
No other functional dependence on $\phi(\bqt)$ is allowed by the gluon fusion
resummation formula.

A general remark about the azimuthal dependence is 
required.
Since we are dealing with collisions of spin unpolarized hadrons,
the corresponding cross sections are invariant under azimuthal rotations
in the transverse plane. Therefore, the multidifferential cross section
$\left[ d\sigma_F \right]^{(g-{\rm fus.})}$ cannot depend on the absolute value 
of $\phi(\bqt)$. The azimuthal dependence of 
$\left[ d\sigma_F \right]^{(g-{\rm fus.})}$ can only appear through
final-state azimuthal {\em correlations}, namely, through functions of relative
azimuthal angles $\Delta\phi_i$, such as, for instance, 
$\Delta\phi_i=\phi(\bqt)-\phi(\bqt_i)$. Here, $\phi(\bqt_i)$ denotes the
azimuthal angle of one of the particles in the produced final-state system $F$
(see Eq.~(\ref{process})). Note that these azimuthal correlations are consistent
with our previous conclusions about the functional dependence on $\phi(\bqt)$;
indeed, we have 
$\cos(2 \Delta\phi_i)= \cos\bigl(2\phi(\bqt)\bigr) \cos\bigl(2\phi(\bqt_i)\bigr)
+ \sin\bigl(2\phi(\bqt)\bigr) \sin\bigl(2\phi(\bqt_i)\bigr)$,
and analogous relations apply to $\sin(2 \Delta\phi_i), \cos(4 \Delta\phi_i)$
and $\sin(4 \Delta\phi_i)$. According to the general notation that we have used
in this paper, the $\phi(\bqt_i)$ dependence of the multidifferential cross
section in Eq.~(\ref{diffxs}) (or, Eq.~(\ref{singsig})) is introduced through the
final-state kinematical variables generically denoted by
${\bf \Omega} = \{\Omega_A, \Omega_B, \cdots\}$. In Eqs.~(\ref{hcg}), 
(\ref{hgc}) and (\ref{hGG}), the dependence on ${\bf \Omega}$
(e.g., on $\phi(\bqt_i)$)
of $H^F(x_1p_1, x_2p_2; {\bf\Omega}; \as(M^2))$
combines with the explicit dependence on 
$\phi(\bqt)$ to produce the final-state azimuthal correlations.
The functional form of the azimuthal correlations is determined by the
$\phi(\bqt)$ dependence of the hard-scattering factors 
$\left[ H^F(\phi(\bqt)) \right]_{I}$, through perturbative coefficients 
(we recall that the Lorentz tensor
$H_{\mu_1 \,\nu_1, \mu_2 \,\nu_2 }^F$, and the equivalent helicity tensor 
$H_{(\lambda_1, \,h_1), (\lambda_2, \,h_2)}^F$, 
are power series functions of $\as$)
that are
process dependent and observable dependent (i.e., they depend on the physical
observable that is specified by the multidifferential cross section
$d\sigma_F/d{\bf \Omega}$). In particular, if the multidifferential cross 
section is insensitive to the azimuthal angles $\phi(\bqt_i)$
(see, e.g., the simple case of inclusive production of the SM Higgs boson),
the corresponding hard-scattering factors $\left[ H^F(\phi(\bqt)) \right]_{I}$
vanish order-by-order in perturbation theory.

The structure of the gluon fusion resummation formula (\ref{qtycrossphi})
is much richer than the structure of the corresponding `\naive' (i.e.,
extrapolated from $q{\bar q}$ annihilation) formula in Eq.~(\ref{qtycross}).
The additional structure is due to the helicity-flip contributions. These
contributions are perturbatively driven by the gluonic coefficient function
$G_{g\,a}(z;\as)$, and they lead to several hard-scattering factors: the factor
$H^F_G$ (see Eq.~(\ref{hGhel})), which contributes to the term
$\left[ d\sigma_F \right]_{\phi}^{(g-{\rm fus.})}$,
and the factors $\left[ H^F(\phi(\bqt)) \right]_{I}$.
We briefly sketch the small-$\qt$ singular behaviour produced by these various
terms in QCD perturbation theory. To this purpose, we expand the integrand of
the resummation formula in powers of $\as=\as(\mu_R^2)$ (with $\mu_R \sim M$),
and we explicitly perform the Bessel transformations from $b$ space to $\qt$
space (technical details on this procedure, and explicit perturbative formulae
can be found, for instance, in Ref.~[\ref{Bozzi:2005wk}]).

The single helicity-flip terms first contribute at the relative order $\as$.
At this perturbative order, they lead to a partonic cross section contribution
that is proportional to
\beq
\label{sflo}
\f{\as}{\pi} \;\delta(1-z_1) \;\delta_{g\,a_1} 
\;G_{g \,a_2}^{(1)}(z_2) \;\left[ H^F(\phi(\bqt)) \right]_{C_1G_2}
\;\left[\f{1}{\qt^2}\right]_+ \;\;,
\eeq
and to an analogous contribution that is obtained by the exchange $1
\leftrightarrow 2$ of the 
subscripts.
Since at this order there is a leading logarithmic term of the type
$\left[\f{1}{\qt^2}\ln (M^2/\qt^2)\right]_+$ (this term is due to the `\naive'
contributions, with no helicity flips), the term in the expression 
(\ref{sflo}) represents a next-to-leading logarithmic effect.
The double helicity-flip term first contributes at the relative order $\as^2$;
it produces a partonic cross section contribution that is proportional to
\beq
\label{dflo}
\left( \f{\as}{\pi} \right)^2 \;G_{g \,a_1}^{(1)}(z_1) 
\;G_{g \,a_2}^{(1)}(z_2)
\left( 
 2 \,\left[ H^F(\phi(\bqt)) \right]_{GG}
\left[\f{1}{\qt^2}\right]_+ + H^F_{G} \;\delta(\qt^2)\;
\right) \;\;,
\eeq
where the functions $G_{g \,a}^{(1)}(z)$ are given in Eqs.~(\ref{gg1}) and 
(\ref{gq1}).

As we have discussed in Sect.~\ref{sec:hel}, our gluon fusion resummation
formula produces differences with respect to the diphoton results presented in
Ref.~[\ref{Nadolsky:2007ba}]. These differences are evident starting from
contributions at the relative order $\as^2$. To explicitly point out the 
${\cal O}(\as^2)$ differences, we can rewrite the structure of Eq.~(38)
in Ref.~[\ref{Nadolsky:2007ba}] by using our notation: this gives an expression
that is proportional to
\beq
\label{dflonad}
\left( \f{\as}{\pi} \right)^2 \;G_{g \,g}^{(1)}(z_1) 
\;G_{g \,g}^{(1)}(z_2) \;\delta_{g\,a_1} \,\delta_{g\,a_2}
\;2 \;\Bigl( 
\,\left[ H^F(\phi(\bqt)) \right]_{GG} + H^F_{G} \,\Bigr)
\;\left[\f{1}{\qt^2}\right]_+ 
\;\;\;,
\eeq
where the sum of hard-scattering factors in the round bracket
originates from Eq.~(30) of Ref.~[\ref{Nadolsky:2007ba}].
The $\qt$ dependence of the expression (\ref{dflonad}) differs from that
of our expression (\ref{dflo}), and the difference is not removed by setting
$G_{g \,q}^{(1)}(z)=G_{g \,{\bar q}}^{(1)}(z)=0$
(the difference between the ${\cal O}(\as)$ expressions in Eq.~(33) 
of Ref.~[\ref{Nadolsky:2007ba}] and in our Eq.~(\ref{sflo}) disappears by
forcing $G_{g \,q}^{(1)}(z)$ and $G_{g \,{\bar q}}^{(1)}(z)$ to vanish).

At higher perturbative orders, the small-$\qt$ singular behaviour in the
expressions (\ref{sflo}) and (\ref{dflo}) is enhanced by powers of 
$\ln (M^2/\qt^2)$. The dominant logarithmic enhancement is produced by the 
gluon form factor $S_g(M,b)$, which appear in 
$\left[ d\sigma_F \right]_{\phi}^{(g-{\rm fus.})}$
(see Eq.~(\ref{qtycrossggphi})) and in each cross section contribution 
$\left[ d\sigma_F \right]_{I}$
(see Eq.~(\ref{qtycrossI})). As is well known, at the relative order 
$\as^n$, the customary (i.e., with no helicity-flip contributions)
{\em leading} logarithmic terms have the following structure:
\beq
\label{nfho}
H^F_{g} 
\;\as^n \;\left( \;
\left[ \,\f{1}{\qt^2} \,\ln^{2n-1}\left(\f{M^2}{\qt^2}\right)\right]_+
\;+ \;\dots \right) \;, \quad \quad \quad \quad \;\;\;\;\;\quad
\quad\;\;\; n \geq 1 \;.
\eeq
At this order, the gluon fusion helicity-flip contributions produce the
following logarithmic behaviour:
\beq
\label{sfho}
\left[ H^F(\phi(\bqt)) \right]_{C_1G_2}
\;\as^n \;\left( \;
\left[ \,\f{1}{\qt^2} \,\ln^{2n-2}\left(\f{M^2}{\qt^2}\right)\right]_+
\;+ \;\dots \right) \;, \quad \quad n \geq 2 \;,
\eeq
\beq
\left[ H^F(\phi(\bqt)) \right]_{GG}
\;\as^n \;\left( \;
\left[ \,\f{1}{\qt^2} \,\ln^{2n-4}\left(\f{M^2}{\qt^2}\right)\right]_+
\;+ \;\dots \right) \;, \quad \quad \;\; n \geq 3 \;,
\eeq
\beq
\label{dfho}
H^F_{G} \;
\as^n \;\left( \;
\left[ \,\f{1}{\qt^2} \,\ln^{2n-5}\left(\f{M^2}{\qt^2}\right)\right]_+
\;+ \;\dots \right) \;, \quad \quad \quad \quad \;\;\;\quad \quad
\;\;\;\;\;
n \geq 3 \;.
\eeq
The dots in the round brackets of Eqs.~(\ref{nfho})--(\ref{dfho})
stand for subdominant terms in each corresponding expression.
The comparison between the expressions (\ref{nfho}) and (\ref{sfho})
shows that the single helicity-flip terms produce {\em next-to-leading}
logarithmic contributions at each perturbative order. The double helicity-flip
terms lead to subdominant logarithmic contributions.

The singular $\qt$ behaviour that is observed order-by-order in the
QCD perturbative expansion is cured by $\qt$ resummation.
We recall that the resummed gluon form factor $S_g(M,b)$
(and, analogously, the quark form factor $S_q(M,b)$ in the 
$q{\bar q}$ annihilation channel) provides the integration over $b$ in
Eqs.~(\ref{qtycrossggphi}) and (\ref{qtycrossI})
with a strong damping factor in the large-$b$ region (roughly speaking,
in the region where $b \gtap {\cal O}(1/M)$).
This damping effect 
(the simple resummation of the leading double-logarithmic terms, 
$\as^n \ln^{2n}(b^2M^2)$, in $b$ space is sufficient to highlight the effect
[\ref{Parisi:1979se}]~)
eventually leads to resummed perturbative predictions for
the $\bqt$ cross section that are physically well-behaved in the small-$\qt$
region. In particular, the qualitative behaviour of the resummed
$\bqt$ cross section (\ref{qtycrossphi}) at very low values of $\qt$
can be examined by performing the limit $\qt \to 0$ of 
Eqs.~(\ref{qtycrossggphi}) and (\ref{qtycrossI}). In this limit, we can write:
\beq
\label{asaver}
\left[ d\sigma_F \right]_{\phi}^{(g-{\rm fus.})} 
\sim \left[ d\sigma_F \right]^{(q{\bar q}-{\rm ann.})}\sim \;{\rm const.} \;\;,
\eeq
\beq
\label{assf}
\left[ d\sigma_F \right]_{C_1G_2} \sim \left[ d\sigma_F \right]_{G_1C_2}
\sim \;\qt^{\,2} \;\;,
\eeq
\beq
\label{asdf}
\left[ d\sigma_F \right]_{GG} \sim \;\qt^{\,4} \;\;.
\eeq
The constant behaviour in Eq.~(\ref{asaver}) is just a result of 
Ref.~[\ref{Parisi:1979se}]. It follows [\ref{Parisi:1979se}] from a simple
reasoning
(and a minor modelling of the Sudakov form factor at very large values of $b$,
$b \sim {\cal O}(1/\Lambda_{QCD})\,$).
In few words, the reasoning amounts to the observation
that the presence of the resummed Sudakov form factor justifies
the use of the low-$\qt$ approximation $J_0(b\qt) \sim 1$ to extract the behaviour
of the resummation formula (\ref{qtycrossggphi}) at $\qt \sim 0$.
The behaviour in Eqs.~(\ref{assf}) and (\ref{asdf}) follows from the same
reasoning.
We simply note that, in the case of the helicity-flip components 
$\left[ d\sigma_F \right]_{I}$ of the $\bqt$ cross section, the resummation
formula (\ref{qtycrossI}) involves higher-order Bessel functions. Thus, we have
just used the low-$\qt$ approximation $J_2(b\qt) \sim b^2 \,\qt^{\,2}$
for the single helicity-flip components (see Eqs.~(\ref{jcg}) and (\ref{jgc}))
and the corresponding approximation $J_4(b\qt) \sim b^4 \,\qt^{\,4}$
for the double helicity-flip component (see Eqs.~(\ref{jgg})).

\section{Summary}

\label{sec:sum}

Considering the hard-scattering production of high-mass systems in 
hadron--hadron collisions, in this paper we have examined the corresponding
transverse-momentum cross sections at small values of $\bqt$.
We have presented a study of the contributions that
are logarithmically enhanced order-by-order in QCD perturbation theory.
The enhanced contributions have the form of singular $\qt$-distributions
of the type 
$\left[\f{1}{\qt^2}\ln^n (M^2/\qt^2)\right]_+$. The all-order analysis and the
perturbative
resummation of these terms, in processes (such as the DY process) that are  
controlled by the $q{\bar q}$ annihilation channel, is a classical QCD result.
We have pointed out that this result does not suffice for the treatment of
processes (such as SM Higgs boson production) that are controlled by
(or, simply, receive contributions from) the gluon fusion channel.
The difference between the $q{\bar q}$ annihilation and gluon fusion channels
originates from correlations that are intrinsically related to the collinear
`evolution' of the colliding hadrons into gluon partonic states.

We briefly summarize the main features of our general results on $\qt$
resummation in gluon fusion processes.
The gluon fusion resummation formula for generic $\bqt$ cross sections
is presented in Eqs.~(\ref{qtycrossgg}), (\ref{whatgg}) and (\ref{cggten}).
The resummation formula controls all the singular (and logarithmically-enhanced)
perturbative contributions to the $\bqt$ cross section in the small-$\qt$
region. Gluon collinear correlations produce new (with respect to $q{\bar q}$
annihilation) structures from the perturbative evolution of the parton 
densities of the colliding hadrons.
The additional structure (see Eq.~(\ref{cggten})) enters $\qt$ resummation
through the factorization formula (\ref{whatgg}).
The terms due to collinear correlations lead, in general, to
next-to-leading logarithmic contributions to the $\bqt$ cross section
(the leading logarithmic contributions still come from soft-radiation effects
included in the customary Sudakov form factor).
Gluon collinear correlations are directly related to helicity-flip phenomena in
the gluon fusion hard-scattering subprocess (see Eq.~(\ref{chelCG})).
These spin correlations originate from a quantum-mechanical interference:
the flip occurs between the gluon helicity states in the scattering amplitude
and in the complex-conjugate amplitude.
The helicity-flip phenomenon due to gluon collinear correlations
leads to definite correlations between the azimuthal angles
of the particles in the high-mass system that is produced 
by the gluon fusion mechanism. These azimuthal-correlation effects 
accompany the dominant (singular) $\qt$ behaviour of the perturbative cross 
section in the small-$\qt$ region (azimuthal correlations with a similar 
$\qt$ behaviour are absent if the high-mass system
is produced by $q{\bar q}$ annihilation).
The functional form of the azimuthal correlations is fully specified, at
arbitrary perturbative orders, by the gluon fusion resummation
formula (see Eq.~(\ref{qtycrossphi}) and Eqs.~(\ref{hcgl})--(\ref{hGG})).
The double helicity-flip component of the $\bqt$ cross section
is characterized by a coherent interference between the spin-flipping
gluons from the two colliding hadrons (see Eq.~(\ref{fbGGhel})). As
a consequence of this interference, the double helicity-flip contribution
produces two distinct terms with a different $\qt$ behavior: a term with
azimuthal correlations (the last term on the right-hand side of 
Eq.~(\ref{qtycrossphi})) and a term, with no azimuthal correlations, that also
contributes to azimuthally-averaged cross sections 
(see Eqs.~(\ref{qtycrossggphi}) and (\ref{whatggphi})). Gluon collinear
correlations thus imply that the differences in the structure of $\qt$
resummation between the gluon fusion and $q{\bar q}$ annihilation channels
persist even after having performed the integration over the azimuthal angle of
the transverse-momentum vector.

An interesting and relevant issue regards the extension of $\qt$ resummation
to processes whose final-state system $F$ contains strongly-interacting 
particles (partons) such as, for instance, high-$p_\perp$ hadrons and, more
generally, jets. The general extension to this type of final-state systems 
(which have not been considered in this paper) is still lacking.
It requires a proper treatment of soft radiation [\ref{softmultiparton}]
in multiparton hard scattering, namely, the hard scattering of the two colliding
partons {\em and} the final-state QCD partons in the system $F$.
The main features of collinear radiation from the two colliding partons are not
affected by the presence of the additional hard partons in the final state. 
Therefore the structure of the gluon collinear correlations that we have found and
documented in this paper is relevant to any extensions of $\qt$ resummation. 
In particular, the convolution of the parton densities with the perturbative
gluonic tensor in Eqs.~(\ref{cggten}) or (\ref{chelCG})
is expected to appear in the $\qt$ resummation formulae for the production of
final-state systems that contain colour-charged partons.
Note that these systems can be produced by $q{\bar q}$ annihilation
(or, generally, $q{\bar q}$ and $qq$ scattering), gluon fusion (or, generally,
$gg$ scattering) and $gq({\bar q})$ scattering subprocesses on equal footing.
Therefore, there is no escape from (almost) ubiquitous collinear correlations 
due to initial-state gluon hard scattering.

\section*{References}

\begin{enumerate}

\item \label{Dokshitzer:hw}
Y.~L.~Dokshitzer, D.~Diakonov and S.~I.~Troian,
Phys.\ Lett.\  B {\bf 79} (1978) 269,
Phys.\ Rep.\  {\bf 58} (1980) 269.

\item \label{Parisi:1979se}
G.~Parisi and R.~Petronzio,
Nucl.\ Phys.\ B {\bf 154} (1979) 427.

\item \label{Curci:1979bg}
G.~Curci, M.~Greco and Y.~Srivastava,
Nucl.\ Phys.\ B {\bf 159} (1979) 451.

\item \label{Collins:1981uk}
J.~C.~Collins and D.~E.~Soper,
Nucl.\ Phys.\ B {\bf 193} (1981) 381
[Erratum-ibid.\ B {\bf 213} (1983) 545].

\item \label{Collins:va}
J.~C.~Collins and D.~E.~Soper,
Nucl.\ Phys.\ B {\bf 197} (1982) 446.

\item \label{Kodaira:1981nh}
J.~Kodaira and L.~Trentadue,
Phys.\ Lett.\ B {\bf 112} (1982) 66,
report SLAC-PUB-2934 (1982),
Phys.\ Lett.\ B {\bf 123} (1983) 335.

\item \label{Altarelli:1984pt}
G.~Altarelli, R.~K.~Ellis, M.~Greco and G.~Martinelli,
Nucl.\ Phys.\ B {\bf 246} (1984) 12.

\item \label{Davies:1984hs}
C.~T.~H.~Davies and W.~J.~Stirling,
Nucl.\ Phys.\  B {\bf 244} (1984) 337.

\item \label{Davies:1984sp}
  C.~T.~H.~Davies, B.~R.~Webber and W.~J.~Stirling,
  Nucl.\ Phys.\  B {\bf 256} (1985) 413.

\item \label{Collins:1984kg}
J.~C.~Collins, D.~E.~Soper and G.~Sterman,
Nucl.\ Phys.\ B {\bf 250} (1985) 199.

\item \label{Catani:vd}
S.~Catani, E.~D'Emilio and L.~Trentadue,
Phys.\ Lett.\ B {\bf 211} (1988) 335.


\item \label{deFlorian:2000pr}
D.~de Florian and M.~Grazzini,
Phys.\ Rev.\ Lett.\ {\bf 85} (2000) 4678,
Nucl.\ Phys.\ B {\bf 616} (2001) 247.

\item \label{Catani:2007vq}
  S.~Catani and M.~Grazzini,
  Phys.\ Rev.\ Lett.\  {\bf 98} (2007) 222002.

\item \label{Catani:2009sm}
  S.~Catani, L.~Cieri, G.~Ferrera, D.~de Florian and M.~Grazzini,
  Phys.\ Rev.\ Lett.\  {\bf 103} (2009) 082001.

\item \label{Becher:2010tm}
  T.~Becher and M.~Neubert,
  report HD-THEP-10-13 
  (arXiv:1007.4005 [hep-ph]).


\item \label{Catani:2000vq}
S.~Catani, D.~de Florian and M.~Grazzini,
Nucl.\ Phys.\ B {\bf 596} (2001) 299.



\item \label{Balazs:2000wv}
  C.~Balazs and C.~P.~Yuan,
  Phys.\ Lett.\  B {\bf 478} (2000) 192;
  Q.~H.~Cao, C.~R.~Chen, C.~Schmidt and C.~P.~Yuan,
  report ANL-HEP-PR-09-20
  (arXiv:0909.2305 [hep-ph]).

\item \label{Berger:2002ut}
  E.~L.~Berger and J.~w.~Qiu,
  Phys.\ Rev.\  D {\bf 67} (2003) 034026,
  Phys.\ Rev.\ Lett.\  {\bf 91} (2003) 222003.

\item \label{Kulesza:2003wn}
  A.~Kulesza, G.~F.~Sterman and W.~Vogelsang,
  Phys.\ Rev.\  D {\bf 69} (2004) 014012.

\item \label{Bozzi:2005wk}
  G.~Bozzi, S.~Catani, D.~de Florian and M.~Grazzini,
  Nucl.\ Phys.\  B {\bf 737} (2006) 73.

\item \label{Bozzi:2007pn}
  G.~Bozzi, S.~Catani, D.~de Florian and M.~Grazzini,
  Nucl.\ Phys.\  B {\bf 791} (2008) 1.


\item \label{Balazs:2006cc}
  C.~Balazs, E.~L.~Berger, P.~M.~Nadolsky and C.~P.~Yuan,
  Phys.\ Lett.\  B {\bf 637} (2006) 235.


\item \label{ZZ}
C.~Balazs and C.~P.~Yuan,
Phys.\ Rev.\  D {\bf 59} (1999) 114007
[Erratum-ibid.\  D {\bf 63} (2001) 059902];
R.~Frederix and M.~Grazzini,
Phys.\ Lett.\  B {\bf 662} (2008) 353.


\item \label{Grazzini:2005vw}
M.~Grazzini,
JHEP {\bf 0601} (2006) 095.


\item \label{Bozzi:2006fw}
  G.~Bozzi, B.~Fuks and M.~Klasen,
  Phys.\ Rev.\  D {\bf 74} (2006) 015001.

\item \label{Dreiner:2006sv}
  H.~K.~Dreiner, S.~Grab, M.~Kramer and M.~K.~Trenkel,
  Phys.\ Rev.\  D {\bf 75} (2007) 035003.


\item \label{Nadolsky:2007ba}
  P.~M.~Nadolsky, C.~Balazs, E.~L.~Berger and C.~P.~Yuan,
  Phys.\ Rev.\  D {\bf 76} (2007) 013008.

\item \label{Mantry:2009qz}
  S.~Mantry and F.~Petriello,
  Phys.\ Rev.\  D {\bf 81} (2010) 093007.


\item \label{CGinprep}
 S.~Catani and M.~Grazzini, in preparation.

\item \label{Vogt:2004mw}
  S.~Moch, J.~A.~M.~Vermaseren and A.~Vogt,
  Nucl.\ Phys.\  B {\bf 688} (2004) 101,
  Nucl.\ Phys.\  B {\bf 691} (2004) 129.

\item \label{Kauffman:1991cx}
  R.~P.~Kauffman,
  Phys.\ Rev.\  D {\bf 45} (1992) 1512.

\item \label{Bozzi:2010xn}
  G.~Bozzi, S.~Catani, G.~Ferrera, D.~de Florian and M.~Grazzini,
  arXiv:1007.2351 [hep-ph].

\item \label{Catani:1996vz}
  S.~Catani and M.~H.~Seymour,
  Nucl.\ Phys.\  B {\bf 485} (1997) 291
  [Erratum-ibid.\  B {\bf 510} (1998) 503].

\item \label{Mantry:2010mk}
  S.~Mantry and F.~Petriello,
  arXiv:1007.3773 [hep-ph].

\item \label{Balazs:2007hr}
  C.~Balazs, E.~L.~Berger, P.~M.~Nadolsky and C.~P.~Yuan,
  Phys.\ Rev.\  D {\bf 76} (2007) 013009.

\item \label{Collins:1977iv}
  J.~C.~Collins, D.~E.~Soper,
  Phys.\ Rev.\  {\bf D16 } (1977)  2219.

\item \label{softmultiparton}
  N.~Kidonakis, G.~Oderda and G.~F.~Sterman,
  Nucl.\ Phys.\  B {\bf 531} (1998) 365;
  R.~Bonciani, S.~Catani, M.~L.~Mangano and P.~Nason,
  Phys.\ Lett.\  B {\bf 575} (2003) 268.

\end{enumerate}

\end{document}